\PassOptionsToPackage{table,xcdraw}{xcolor}
\documentclass[sigconf]{acmart}

\copyrightyear{2024}
\acmYear{2024}
\setcopyright{rightsretained}
\acmConference[ICSE '24]{2024 IEEE/ACM 46th International Conference on Software Engineering}{April 14--20, 2024}{Lisbon, Portugal} \acmBooktitle{2024 IEEE/ACM 46th International Conference on Software Engineering (ICSE '24), April 14--20, 2024, Lisbon, Portugal}\acmDOI{10.1145/3597503.3639584}
   
\acmISBN{979-8-4007-0217-4/24/04}

\usepackage{graphicx} %
\usepackage{multirow}
\usepackage{url}
\usepackage{subfig}
\usepackage{tabularx}
\usepackage{soul,color}
\usepackage{algorithmic}
\usepackage[ruled,vlined,linesnumbered]{algorithm2e}
\usepackage{amsthm}
\usepackage{pifont}
\usepackage{bbding}
\usepackage{enumitem}
\usepackage{dblfloatfix}
\usepackage[nameinlink,noabbrev]{cleveref}
\crefformat{footnote}{\textsuperscript{#2#1#3}}

\newcommand{\approach}{\texttt{DeepSample}}

\title{DeepSample: DNN sampling-based testing for operational accuracy assessment}

\author{Antonio Guerriero, Roberto Pietrantuono, Stefano Russo}
\affiliation{
    \institution{\textit{DIETI, Universit\`a degli Studi di Napoli Federico II}}
  \city{Via Claudio 21, 80125 - Napoli}
  \country{Italy}}
  \email{{antonio.guerriero, roberto.pietrantuono, stefano.russo}@unina.it}

\date{January 2024}

\begin{document}

\begin{abstract}
Deep Neural Networks (DNN) are core components for classification and regression tasks of many software systems.
Companies incur in high costs for testing DNN 
with datasets representative of the inputs 
expected 
in operation, 
as these need to be manually labelled. 
The challenge is to select a representative set of test inputs as small as possible to reduce the labelling 
cost, while sufficing to yield unbiased high-confidence estimates of the expected DNN accuracy. 
At the same time, testers are interested in exposing as many DNN mispredictions as possible to improve the DNN, ending up in the need for techniques pursuing a threefold aim: small dataset size, trustworthy estimates, mispredictions exposure.  

This study presents \texttt{DeepSample}, a family of DNN testing techniques for cost-effective accuracy assessment based on probabilistic sampling. 
We investigate whether, to what extent, and under which conditions probabilistic sampling can help to tackle the outlined challenge. 
We implement five new sampling-based testing techniques, and perform a comprehensive comparison of such techniques and of three further state-of-the-art techniques for both DNN classification and regression tasks. Results serve as guidance for best use of sampling-based testing for faithful and high-confidence estimates of DNN accuracy in operation at low cost. 
\end{abstract}



\maketitle

\section{Introduction}
A countless number of software systems today rely on Deep Neural Networks (DNN) predictions. %
Before release, engineers need to test the DNN to estimate their accuracy (i.e., probability of not having mispredictions). This allows to establish a release criterion and to correct or tune the DNN 
until the criterion is met. 

The reference scenario is the following: a DNN model meant to operate in a target context is trained with a \textit{training dataset}. The goal of the tester %
is to select a small yet representative subset of (unlabelled) inputs from an \textit{operational dataset}, to use as test cases to estimate the DNN accuracy \cite{Li19}. Their manual labelling has a high cost. 
The challenge is to build a small test set 
able to provide an unbiased, high-confidence estimate of the DNN accuracy.
At the same time, testers are interested in exposing DNN mispredictions, since 
they 
are input to DNN debugging and re-training \cite{Guerriero21}. The goal thus becomes threefold: build a \textit{small dataset}, able to faithfully \textit{estimate DNN accuracy}, and with a good ability to \textit{expose mispredictions}.  

Inspired by \textit{operational testing}, a known practice in software reliability engineering \cite{Currit1986, mills1, mills4, mills3, SRET1996}, researchers proposed probabilistic sampling to test DNN. 
The basic scheme is simple random sampling (SRS).
Li \textit{et al.} proposed %
a sampling scheme aimed at minimizing cross-entropy between the selected tests and the operational dataset \cite{Li19}. %
Guerriero \textit{et al.} \cite{Guerriero21} leveraged adaptive sampling \cite{Thompson} to propose DeepEST, whose objective is to 
expose many DNN mispredictions while yielding good accuracy estimates. %
These techniques borrow basic concepts from sampling theory to derive algorithms working well for specific goals or contexts -- for instance, CES and DeepEST outperform each other in their respective objectives (lower-variance estimate the former, better 
failure exposure the latter). 
However, better trade-offs can be achieved by exploiting advanced strategies from statistical sampling, e.g., by properly using the information available to drive the sampling process. %

This work aims to give
a high level view of sampling-based DNN testing to highlight what are the main knobs to tailor a technique according to the needs and improve performance, exploiting 
advanced sampling theory concepts 
besides the basic ones 
(e.g., auxiliary variables, unequal sampling, without-replacement schemes, stratification).
To this aim:

\begin{itemize}[leftmargin=*]
\item We propose \texttt{DeepSample}, a family of 
sampling-based DNN testing techniques differing from each other in the sampling strategy, in the auxiliary information used for sampling and for partitioning, and in the estimation process. 
 
The framework includes %
five new testing techniques,  
each implemented in three variants depending on the auxiliary information used to drive sampling. %

\item We present a comprehensive comparison of the new techniques
and of three existing ones, SRS, CES, DeepEST, to
evaluate their ability to 
assess DNN accuracy 
and select failing examples. 
The evaluation 
is conducted on classification and regression
tasks, under 5 testing budgets, 3 datasets, with 3
models per dataset for classification, and 1 dataset and 2 models 
for regression.\footnote{The replication package is at: \url{https://github.com/dessertlab/DeepSample.git}.\label{repo}}
\end{itemize}

The new algorithms turn out to outperform the existing ones in almost all the contexts. %
Overall, the results allow to draw guidelines for practitioners and researchers - on relevant factors like if and which auxiliary information to use and how to use it - for sampling-based DNN testing for high accuracy, high confidence estimates at low cost and with good mispredictions exposure ability.

\section{Related work}
\label{sect:Related}
Probabilistic sampling is used in \textit{operational testing} (OT) to estimate the expected reliability of a software system after release.  
In OT, test suites are built by selecting or generating tests according to the expected \textit{operational profile}, a probabilistic characterization of the expected usage.  %
OT was central in Cleanroom software engineering \cite{mills1, Currit1986, mills3, mills4}
and in the Software Reliability Engineering Test process \cite{SRET1996}.%

Over the years, researchers proposed better sampling strategies to improve estimates or lower their cost. 
Cai \textit{et al.} \cite{CaiJSS, cai7, cai5} developed \textit{Adaptive Testing}, still based on the operational profile, but with an adaptive selection of test cases from partitions. Adaptive Testing with Gradient Descent\footnote{
At each step, the partition selected to draw the next test is the one that yields the greatest descent (i.e., negative gradient) of the variance of the reliability estimator.}
\cite{CaiTSE} is one of the techniques considered in this study. Stratified sampling too has been used for reliability assessment \cite{Podgurski1999, omri2014}. Later, Pietrantuono \textit{et al.} \cite{RELAI, QRS2018} stressed the use of unequal probability sampling to improve efficiency, formalizing several sampling schemes to this aim \cite{ISSRE2016}.

Li \textit{et al.} \cite{Li19} first proposed sampling for DNN operational accuracy assessment in the CES (Cross-Entropy Sampling) technique. 
Like OT, CES aims to select a small yet representative sample, by minimizing the cross-entropy between the selected and the operational dataset. A sample is expected to contain the same proportion of failing examples as in the operational dataset. 
Guerriero \textit{et al.} \cite{Guerriero21} observed that the mere imitation of operational inputs may be inefficient, especially for accurate DNN, 
as much effort is wasted to label correctly classified inputs. 

 They propose DeepEST, exploiting an \textit{adaptive sampling} algorithm for rare populations \cite{Thompson} to spot the %
 more failing examples, hence spending effort to label examples useful for improvement besides assessment. 
 The disproportional selection is balanced by an estimator that preserves unbiasedness.

A further technique is PACE (Practical accuracy estimation) \cite{Chen2020-2}, a heuristic method that uses clustering to partition tests into groups, and then uses adaptive random selection of test inputs representative of the clusters. 
Zhou \textit{et al.} proposed 
DeepReduce \cite{Zhou2020}, a two-stage heuristic method exploiting neuron coverage to select a subset of inputs, then using the Kullback-Leibler Divergence
to drive the second-stage selection. 
These techniques are however not based on probabilistic sampling like those compared in this work, and they do not guarantee unbiasedness and convergence.

\section{Sampling-based Testing} %
\label{sect:DeepSample}
\subsection{Formulation}
\label{sect:DeepSample_form}
\begin{itemize}[leftmargin=*]%
    \item $M$ is the DNN model under test; 
    \item $D = \{d_1, \dots, d_N\}$ is the \textit{operational dataset}, an arbitrarily large set of examples with unknown labels, which are possibly given as input to the model $M$ in the operational phase. Its size is $N=|D|$;
    \item $T \in D = \{t_1, \dots, t_n\}$ is the subset of examples to select from $D$ and to be  labelled. This set is used for estimating DNN  accuracy, and can also be used to enlarge the training set and improve the DNN performance in new releases. Its size is $n=|T|$ $\ll N$.
    When an example $t_i$ is submitted to the DNN, a human oracle assigns the expected output to $t_i$, and then 
    compares it with the actual output. In classification tasks, this gives a binary outcome $z_i$ (whether actual and expected labels match or not). In regression tasks, the comparison gives an offset $\delta_i$, which is the absolute difference between the true ($r_i$) and predicted ($\hat{r}_i$) values -- considering this a failure or not depends on the tolerable threshold. For our purposes, it suffices to focus on the value of $\delta_i$.

    \item $\theta = Pr(z_i = 1)$, with $i=1,\dots |D|$, is, in classification tasks, the true failure probability on a randomly selected example from the entire operational dataset, and corresponds to the true (unknown) proportion $\theta = \frac{1}{N}\sum_{i=1}^N z_i $. Accuracy is defined as: $\xi = 1-\theta$.  In the case of regression, we look at the mean squared error between the true ($r_i$) and predicted ($\hat{r}_i$) value over the entire operational dataset: $\Delta = \frac{1}{N}\sum_{i=1}^N \delta_i^2$, and $\xi = 1 -  \Delta$. Its estimate is $\hat{\xi}$. %

\end{itemize}

Given a sample size budget $n$, the goal of \approach{} is to select a subset $T$ able of giving an \textit{unbiased} (i.e., such that $\mathbb{E}[\hat{\xi}]=\xi$) estimate of $\xi$ while maximizing the \textit{efficiency} of the estimator (i.e., minimizing the variance of the estimate).\footnote{Minimizing the variance is equivalent to minimize the MSE since the estimators are required to be unbiased. Low variance (or MSE) implies maximizing the confidence.} In addition, the set $T$ is wanted to expose as many failing examples as possible.

\subsection{Overview of DeepSample}

\approach{} is a family of techniques leveraging  prior knowledge available about the operational dataset, 
supposed to be correlated to the variable to estimate (namely, accuracy).  %
Prior information is encoded in what are called \textit{auxiliary variables} \cite{BOOK1}, here denoted as $\chi$; for instance, the confidence value provided by classifiers when predicting a label can be assumed to be (negatively) correlated with the failure probability $\theta$. Clearly,  accuracy and efficiency of estimates depend on the extent to which assumptions hold. 

The \approach{} techniques are characterized by two dimensions:  
\textit{i)} the \textbf{sampling algorithm}, and \textit{ii)} the \textbf{auxiliary variable}.  

The former specifies a \textit{sampling scheme}, namely the sequence of steps required to select the tests $t_i$. 
The latter specifies what is the \textit{auxiliary variable} $\chi$, if used by the sampling scheme (not all auxiliary variables can be used in all the schemes). 

There are two ways of exploiting the auxiliary variables. The first is to partition the dataset into classes that are homogeneous with respect to the auxiliary variable (e.g., similar confidence), similarly to stratification in sampling theory \cite{BOOK1}. If the variable is well correlated with the failure probability $\theta$ (or $\Delta$ for regression), partitions too should be homogeneous with respect to $\theta$ (or $\Delta$). This allows to wisely allocate the number of examples to draw from each partition with the aim to reduce the variance of the estimation. %
The second way is to let the sampling scheme select the examples proportionally to the auxiliary variable's value, so as to get the ones with higher expected failure probability. 
A proper estimator is then needed to correct bias due to this \textit{unequal} selection probability. %

Techniques can be \textit{with} or \textit{without replacement}. The former ones (allowing an example to be selected more times) are associated with simpler estimators - a common choice in literature \cite{Li19} \cite{CaiJSS} \cite{cai7} \cite{cai5} \cite{CaiTSE}; the latter ones are expected to give higher efficiency, though the gain in large populations can be marginal (with-replacement schemes will unlikely select twice the same example). 

The \textbf{estimator} takes the result of submitting the selected sample $T$ to the DNN $M$ (denoted as $z_i$ and $\delta_i$ for classification and regression, respectively) and yields an unbiased estimate of $\xi$ by counterbalancing the disproportional selection (cf. with Sec. \ref{techniques}). %

\subsection{Auxiliary variables}
We consider three auxiliary variables for classification problems, and for regression as well. For classification, they are:
Confidence, Distance-based Surprise Adequacy (DSA), and Likelihood-based Surprise Adequacy (LSA). We opted for these variables based on the literature \cite{Li19, Guerriero21}.
For regression, they are LSA, and two variables based on the reconstruction error of a simple autoencoder (SAE) and of a variational autoencoder (VAE), which have been demonstrated to be effective in detecting  inputs likely to cause failure \cite{Stocco2020}.

Confidence $C_{d_i}$ of an input $d_i$ is the maximum value in the probability vector obtained from the last layer's output of the DNN\footnote{In the case of binary classification with a single neuron, the confidence is the neuron output $o$ when $o\geq0.5$ (e.g., class 1) and $1-o$ when $o<0.5$ (class 0).}; it is for classification problems only.
DSA and LSA, defined by Kim \textit{et al.} \cite{Kim19}, exploit Activation Traces (AT), which are vectors of activation values of neurons belonging to a certain layer. 
DSA is defined as: %
    $DSA_{d_i} = \frac{\sigma_A}{\sigma_B}$, 
where $\sigma_A$ is the Euclidean distance between the ATs of the input $d_i$ (whose predicted class is A) and its nearest neighbour belonging to the same class $A$, $\sigma_B$ is the distance between the ATs of $d_i$ and its nearest neighbour belonging to a different class $B$.\footnote{Note that the computation does  not need the actual labels (but only predicted ones).} It makes sense for classification models only.
LSA uses Kernel Density Estimation (KDE) \cite{wand1994kernel} to estimate the probability density of each activation value, obtaining the surprise of a new input with respect to the estimated density. LSA is a measure of rareness computed as: 
    $LSA_{d_i}= -log(\hat{f}(d_i))$, 
where $\hat{f}(d_i)$ is the KDE applied to the new input $d_i$. LSA is for both classification and regression.

For SAE/VAE-based variables,  we leverage 
the reconstruction error $\epsilon$. %
We used the two best-performing autoencoders implemented by Stocco \textit{et al.} \cite{Stocco2020},  SAE (Simple Autoencoder) with a single hidden layer, and VAE (Variational Autoencoder). We consider autoencoders as single-image reconstructors, computing their outputs for all the operational examples, and then calculating the reconstruction error as: $  \epsilon_{d_i} = \frac{1}{WHC}\sum_{k=1,j=1,c=1}^{W,H,C}(d_i[c][k,j]-d_i'[c][k,j])^2$, where $d_i$ is the original image, $d_i'$ is the reconstructed image, $W$, $H$, and $C$ are width, height, and channels respectively. 
The corresponding auxiliary variables are synthetically called SAE and VAE, meaning the $ \epsilon_{d_i}$ value obtained by SAE and VAE.

All variables are assumed to be correlated to accuracy: lower confidence, higher surprise (DSA, LSA), and higher reconstruction error (SAE, VAE) are expected to be related to higher failure probability.  
To have all positive variables (from which selection probabilities need to be derived), DSA and LSA for classification are \textit{min-max} normalized. For regression, as the min-max normalization affects the distribution of test data, we just shift the values: 
$DSA_{d_i} = DSA_{d_i} + \lceil \left | min(DSA_{d_i}) \right | \rceil$ (the same for $LSA$).
All the above variables are denoted as $\chi_i$
in the following, when there is no need to distinguish them. In the case of confidence, $\chi_i=1-C_{d_i}$ since we assume that confidence is negatively correlated to the accuracy.

\subsection{Testing techniques}
\label{techniques}
The characteristics of the eight compared testing techniques are summarized in Table \ref{summary:table}; their description follows. 

\begin{table}[t]
	\centering
	\caption{
 Compared testing techniques}
	\label{summary:table}
 \vspace{-6pt}
 \resizebox{\columnwidth}{!}{ 
\begin{tabular}{l|c|c|c|c|c|c|c|c}
\toprule
\textbf{Technique} & {SUPS}              & {RHC-S}                      & {SSRS}              & {GBS}               & {2-UPS}  & {SRS}    & {CES}               & {DeepEST}           \\ \hline 

\textbf{Partitioning} & \ding{55} & \ding{55} & \ding{51} & \ding{51} & \ding{51} & \ding{55} & \ding{55} & \ding{55} \\ \hline

\textbf{\begin{tabular}[c]{@{}c@{}}Unequal\\ selection\end{tabular}}  & \ding{51} & \ding{51}  & \ding{55} & \ding{55} & \ding{51}& \ding{55} & \ding{51} & \ding{51}\\ \hline

\textbf{\begin{tabular}[l]{@{}l@{}}Without \\ replacement\end{tabular}} & \ding{55} & \ding{51}   & \ding{51} & \ding{55} & \ding{51}  & \ding{55}& \ding{55} & \ding{51} \\ \bottomrule
\end{tabular}}
\vspace{-9pt}
\end{table}

\subsubsection{Without-partitioning techniques\label{without_partition}}
\paragraph{Simple Random Sampling (\textbf{SRS})} 
SRS with replacement, where all examples have the same probability to be selected, 
is the simplest and baseline technique  \cite{ISSRE2016}\cite{Li19}. %
 For SRS; unbiased estimators of $\theta$ (for classification) and $\Delta$ (regression) are, respectively, the observed proportion and mean squared error over the subset of selected tests: 

\noindent
\begin{minipage}{.45\columnwidth}
\footnotesize{ 
\begin{equation}
  \hat{\theta} = \frac{1}{n} \sum_{i=1}^n z_i
  \label{srs_classification}
\end{equation}
}
\end{minipage}
\hspace{0.5cm}
\begin{minipage}{.45\columnwidth}
\footnotesize{ 
\begin{equation}
  \hat{\Delta} = \frac{1}{n} \sum_{i=1}^n \delta_i^2
  \label{srs_regression}
\end{equation}
}
\end{minipage}%

\paragraph{Simple Unequal Probability Sampling (\textbf{SUPS})}
This scheme leverages auxiliary variables $\chi$ for selecting the examples. %
The selection probability $\pi_i$ for the $i$-th example $t_i$ is obtained by normalizing the auxiliary variable  $\pi_i = \chi_i/\sum_{i=1}^N \chi_i$; this is known as probability-proportional-to-size (PPS) sampling \cite{BOOK1}. 
The selection is with replacement. An unbiased estimator is the sample mean of the observed values re-scaled by the inverse of their selection probability $\pi_i$ and by $N$, known as Hansen-Hurwitz estimator \cite{Hansen1943}:  

 \vspace{3pt}
 \noindent\begin{minipage}{.45\columnwidth}
\footnotesize{
\begin{equation}
  \hat{\theta} = \frac{1}{nN} \sum_{i=1}^n \frac{z_i}{\pi_i}
\end{equation}
}
\end{minipage}%
\hspace{0.5cm}
\begin{minipage}{.45\columnwidth}
\footnotesize{ 
\begin{equation}
  \hat{\Delta} = \frac{1}{nN} \sum_{i=1}^n \frac{\delta_i^2}{{\pi_i}}
\end{equation}
}
\end{minipage}%

Note that this is a generalization of SRS, wherein the selection probability is $\pi_i = 1/N$ for all the examples.  

\paragraph{RHC-Sampling (\textbf{RHC-S})} 
This is another unequal probability selection scheme, but without replacement, and uses the Rao, Hartley, and Cochran (RHC) estimator \cite{RHC}. The scheme is as follows: 
\begin{enumerate}[leftmargin=5.5mm]
\item Given the budget of $n=|T|$ test cases, divide randomly the $N = |D|$ units of the operational dataset into $n$ groups, by selecting $G_1$ inputs with SRS \textit{without replacement} for the first group, then $G_2$ inputs out of the remaining ($N-G_1$) for the second, and so on. This will lead to $n$ groups of size $G_1$, $\dots$, $G_n$ with $\sum_{r=1}^n G_r = N$. The group size is arbitrary, but we select $G_1 = G_2=\dots=G_n = N/n$, as this minimizes the variance.  
\item One test case is then drawn by taking an input $t_i$ in each of these $n$ groups independently and with a PPS sampling according to the above-defined $\pi$ variable. 
\item Denote with $\pi_{i,r}$ the probability associated with the $t_i$-th unit in the $r$-th group, and with $q_r = \sum_{i \in G_r} \pi_{i,r}$ the sum in the $r$-th group. The unbiased estimators are: 

\noindent\begin{minipage}{.43\columnwidth}
\footnotesize{
\begin{equation}
  \hat{\theta} = \frac{1}{N} \sum_{r=1}^n \frac{z_{r}}{ \pi_{r}  / q_r} 
\end{equation}
}
\end{minipage}%
\begin{minipage}{.43\columnwidth}
\footnotesize{ 
\begin{equation}
\hat{\Delta} = \frac{1}{N} \sum_{r=1}^n \frac{\delta^2_{r}}{ \pi_{r}  / q_r} . 
\end{equation}
}
\end{minipage}%

 \end{enumerate}

\paragraph{Cross-entropy Sampling (\textbf{CES})}  
Cross Entropy-based Sampling (CES) was proposed by Li \textit{et al.} \cite{Li19}. 
The CES algorithm builds the sample first selecting randomly an initial set of examples, and then selecting the remaining examples trying to minimize the average cross-entropy between the probability distribution of the $m$-dimensional representation of neurons output computed on the operational dataset and the selected images. The objective is to sample a set of examples as much as possible representative of the operational dataset, namely if it contains the same proportion of mispredictions as the operational dataset.
For CES, the authors demonstrate that the estimator is the same as SRS  (Eq. \ref{srs_classification} and Eq. \ref{srs_regression}). 

\paragraph{Deep neural networks Enhanced Sampler for operational Testing (\textbf{DeepEST})}
Guerriero \textit{et al.} presented DeepEST \cite{Guerriero21}, 
a technique for DNN operational testing with the twofold objective of accuracy estimation and accuracy improvement. 
DeepEST exploits adaptive sampling \cite{Thompson} to select a sample providing a close and efficient estimate \textit{and}, at the same time, including a high number of failing examples. %
The original version of DeepEST works only for classification tasks. We hereafter extend it for regression too, defining the corresponding estimator. 
The auxiliary variable, $\chi$, is used by DeepEST to define a \textit{weight} $w_{i,j}$ between any pair of examples $d_i$ and $d_j$ of the operational dataset, used to explore the example space adaptively. The weight $w_{i,j}$ is the value of $\chi_{d_j}$ if 
 $\chi_{d_i}$ exceeds a threshold (i.e., it means that $t_i$ is in an interesting cluster to explore), 0 otherwise. The thresholds are those of the original paper. The strategy acts as follows: the first input is selected via SRS, then a \textit{weight-based sampling} (WBS) is used with probability \textit{r} to sample the next example (or SRS with probability \textit{1-r}). 
The example $d_i$ is selected at step $k$ with probability $q_{k,t_i}$: 
\begin{equation}
\footnotesize{
q_{k,i}=r \cdot \frac{\sum_{j \in s_k} w_{i,j}}{\sum_{h \notin s_k, t_j \in s_k} w_{h,j}}+(1-r) \cdot \frac{1}{N-n_{s_k}} 
\label{eq1}
}
\end{equation}
\noindent where:
\begin{itemize}
	\item $r$: probability of using WBS;
	\item $s_k$: current sample (all examples selected up to step $k$); 
	\item $w_{i,j}$: weight relating example $d_j$ in $s_k$ to example $d_i$; 
	\item $n_{s_k}$: the size of the current sample $s_k$;
	\item $N$: the size of the operational dataset. 
\end{itemize}

WBS selects an example $d_i$ proportionally to the sum of weights $w_{i,j}$ of already selected examples toward $d_i$. 
We compute the following step-by-step estimators to balance for the adaptive sampling: 

\noindent\begin{minipage}{.45\columnwidth}
\footnotesize{
\begin{equation}
  \hat{\theta} = \frac{1}{n} (z_1 + \frac{1}{N}\sum_{k=2}^n \tilde{\theta}_k)
  \label{RHC_est_1}
\end{equation}
}
\end{minipage}%
\begin{minipage}{.45\columnwidth}
\footnotesize{
\begin{equation}
\hat{\Delta} = \frac{1}{n} (\delta^2_1 + \frac{1}{N}\sum_{k=2}^n \tilde{\Delta}_k)
\label{RHC_est_2}
\end{equation}
}
\end{minipage}%

\noindent where $z_1$ and $\delta^2_1$ are the estimates obtained at step $k=1$ (hence when $n=1$), $\tilde{\theta}_k$ and $\tilde{\Delta}_k$ are the Hansen-Hurwitz estimates at step $k>1$ for the total failures and for the mean-squared error: 

\noindent\begin{minipage}{.5\columnwidth}
\footnotesize{
\begin{equation}
  \tilde{\theta}_k = \sum_{j \in s_{k}} z_j + \frac{z_i}{q_{k,i}} 
\end{equation}
}
\end{minipage}%
\begin{minipage}{.5\columnwidth}
\footnotesize{
\begin{equation}
\tilde{\Delta}_k = \sum_{j \in s_{k}} \frac{\delta^2_j}{k-1} + \frac{\delta^2_i/k}{q_{k,i}} .
\end{equation}
}
\end{minipage}%
\vspace{6pt}

The final estimators (Eq. \ref{RHC_est_1}, \ref{RHC_est_2}) are the sample mean of the step-by-step estimators. For regression, the $k$-th MSE estimate is $\tilde{\Delta}_k$.

\subsubsection{Partition-based techniques} 
Partition-based techniques split the operational dataset into classes to improve sampling. In sampling theory, stratification
 splits the population to have a small expected intra-stratum variance of the variable to estimate $\xi$ 
 and a large inter-strata variance, so as to sample more from partitions with higher variance. %
 Since the true variance of $\xi$ is unknown, stratification can be done on an estimate of such variance (e.g., computed from a preliminary sample) \cite{ISSRE2016}. However, this would require labeling a subset only just for the purpose of estimating the variance and then applying stratification. Another common solution, that we adopt, is to %
 stratify based on auxiliary variables. Although risky (performance depends on the extent to which they are correlated to $\xi$), this requires no prior knowledge about $\xi$.  We used $k-$means clustering \cite{macqueen1967some} on $\chi$, with $k$ set to 10 after a preliminary tuning on 30 random samples from MNIST, with $k= 6, 8, 10, 12$. 

\paragraph{Stratified Simple Random Sampling (\textbf{SSRS})}  
In this scheme, the number of examples to draw from each partition $p$ is computed by the Neyman allocation \cite{BOOK1} applied to  $\chi$, namely proportionally to the standard deviation of the (normalized) $\chi$ values for that partition, and to 
the size of the partition, $N_p$. 
Selection within the partition is without-replacement. The estimators are the weighted sum of the SRS estimates for partitions:

\noindent\begin{minipage}{.45\columnwidth}
\footnotesize{
\begin{equation}
\hat{\theta} = \frac{1}{N} (\sum_{p=1}^P N_p\hat{\theta}_p)
\label{SSRSeq1}
\end{equation}
}
\end{minipage}%
\hspace{0.5cm}
\begin{minipage}{.45\columnwidth}
\footnotesize{
\begin{equation}
\hat{\Delta} = \frac{1}{N} (\sum_{p=1}^P N_p\hat{\Delta}_p)
\label{SSRSeq2}
\end{equation}
}
\end{minipage}%

\noindent where $\hat{\theta}_p$ and $\hat{\Delta}_p$ are the within-partition SRS estimators (Section \ref{without_partition}), $P=k=10$ is the number of partitions.

\paragraph{Gradient-Based Sampling (\textbf{GBS})}
Unlike SSRS, this technique does not initially allocate a sample size for each stratum, but it decides step by step which partition the next example will be drawn from. Inspired by adaptive testing with gradient descent \cite{CaiTSE}, at each step the partition is chosen so as to maximize the reduction of the variance $Var(\hat{\xi})$ of the $\xi$ estimator, by taking the partition with the largest negative gradient: 
 $-\partial Var(\hat{\xi})/\partial n_p$ (ties broken randomly), $n_p$ being the number of examples selected from partition $p$ up to the current step. %
The selection within the partition is then with replacement. The estimators are the same as SSRS (Eq. \ref{SSRSeq1}, \ref{SSRSeq2}). %
Note that the with-replacement SRS, used in GBS, and without-replacement SRS, used in SSRS, have the same mean estimators -- they differ for the variance of these estimators.  

\paragraph{Two-stage Unequal Probability Sampling (\textbf{2-UPS})}
This technique implements a two-stage sampling scheme, where unequal probability sampling is adopted to select the partition (first stage), and SRS without replacement is adopted to select the example from the chosen partition (second stage). 
The selection probability for partition $p$ is proportional to the sum of (normalized) $\chi$ values (denoted as $\pi_i$ as in SUPS and RHC-S) within that partition: 
\begin{equation}
\footnotesize{
\psi_p = \frac{\sum_{i=1}^{N_p} \pi_i}{\sum_{p=1}^P \sum_{i=1}^{N_p} \pi_i} . 
}
\end{equation}

Clearly, selection of partitions is with replacement; $Q_p$ is the number of times partition $p$ is selected. The estimator for this technique is the average over $n$ estimates: %

\noindent\begin{minipage}{.45\columnwidth}
\footnotesize{
\begin{equation}
\hat{\theta} = \frac{1}{Nn} (\sum_{p=1}^P \sum_{i=1}^{Q_p} \frac{z_i N_p}{\psi_p})
\end{equation}
}
\end{minipage}%
\hspace{0.5cm}
\begin{minipage}{.45\columnwidth}
\footnotesize{
\begin{equation}
\hat{\Delta} = \frac{1}{Nn} (\sum_{p=1}^P \sum_{i=1}^{Q_p} \frac{\delta^2_i N_p}{\psi_p})
\end{equation}
}
\end{minipage}%

Inner terms $(z_i N_p)/(\psi_p)$ and $(\delta^2_i N_p)/(\psi_p)$ are Hansen-Hurwitz estimates for the total number of failures and squared errors in partition $p$, respectively. %
 These estimates are summed up over all partitions and divided by the sample size $n$ to get an average total estimate. The division by $N$ gives  $\hat{\theta}$ and $\hat{\Delta}$.

\section{Evaluation}
\label{sect:Evaluation}
\subsection{Research questions and metrics}
\noindent \textbf{RQ1}: \textit{How do the sampling techniques perform in assessing the operational accuracy of DNN models?    }
    \begin{itemize}[leftmargin=*]
        \item \textbf{RQ1.1}: \textit{How do the techniques perform for classification?}
        \item \textbf{RQ1.2}: \textit{How do the techniques perform for regression? }
    \end{itemize}
    Over $R = 30$ repetitions, we measure the root mean squared error (RMSE) between the accuracy estimates $\hat{\xi}$ and the true  accuracy $\xi$ computed on the operational datasets by labeling all the images: 
    \begin{equation}
    \footnotesize{
    RMSE = \sqrt{\frac{\sum_{r=1}^R(\xi-\hat{\xi})^2}{R}}
    }
    \label{rmse}
    \end{equation}
    where $\hat{\xi}$ for classification and regression is computed using $\hat{\theta}$ and $\hat{\Delta}$, respectively. %
    Lower RMSE means higher confidence in the estimate. 
    
\vspace{3pt}
\noindent \textbf{RQ2}: \textit{How do the sampling techniques perform in detecting failing examples?} An issue of some techniques like CES is that they, with reason, try to have in the sample the same proportion of failures as in the operational dataset, to faithfully estimate accuracy (what is called the \textit{imitation bias} \cite{Guerriero21}); but in highly-accurate DNNs, this entails  %
    very few failures exposed, which requires engineers to run further tests to expose failures 
    -- an issue addressed by DeepEST \cite{Guerriero21}. Thus a desirable property is to expose a high number of failures, besides the ability to provide unbiased high-confidence estimates. 
    \begin{itemize}[leftmargin=*]
        \item \textbf{RQ2.1}: Classification task. \textit{How many failures (namely, misclassifications) are exposed by the techniques?} 
        \item \textbf{RQ2.2}: Regression task. \textit{How many examples with an inaccurate prediction are selected by the techniques?}     
        Since in regression we have continuous outputs, we measure the number of examples having a difference between true and predicted output %
        (i.e., the offset: $\delta_i = |r - \hat{r}_i|$) 
        greater than or equal to a given value $y$: $N_{\delta \geq y}$ with $y$ ranging from 0$^\circ$ to 25$^\circ$, with a step of  2.5$^\circ$.\footnote{The output of the DNN for regression is a steering angle degree; a difference greater than 25$^\circ$ is unrealistic, and never occurred in our experiments.} %
    \end{itemize}

    \noindent \textbf{RQ3}: \textit{How does the budgeted sample size affect performance?} The sample size is directly related to the cost of labelling, as it determines the number of examples to be manually labelled. %
    \begin{itemize}[leftmargin=*]
        \item \textbf{RQ3.1}: \textit{How does the size affect the accuracy estimate? }%
        \item \textbf{RQ3.2}:  \textit{How does the size affect the failing examples detection?}%
    \end{itemize}

To answer RQ1 and RQ2, we consider a budget size of $200$, as in  \cite{Li19, Guerriero21}; the total runs are $6,$$600$ [$11$ models $\times$ $30$ repetitions $\times$ ($6$ techniques $\times$ $3$ auxiliary variables + the $2$ techniques CES and SRS not using auxiliary variables)]. For RQ3, with 5 sample size values ($50$, $100$, $200$, $400$, $800$), there are additional $6,$$600$ $\times$ $4 = 26,$$400$ runs, for a total of $33,$$000$ runs. %

\subsection{Subjects}%

The evaluation is on 11 DNN models on popular datasets (Table \ref{tab:models}). For classification we consider 3 models for each of the following 3   datasets: 
MNIST \cite{LeCun2010}, 
CIFAR10 %
and CIFAR100 %
\cite{Krizhevsky09}.
MNIST has $70,$$000$ entries; CIFAR10 has $60,$$000$ entries; both have 10 classes. CIFAR100 also has $60,$$000$ entries, with 100 classes.
For regression, we consider 2 models for the Udacity dataset\footnote{\url{https://github.com/udacity/self-driving-car}.} ($101,$$396$ entries for training,  $5,$$614$ for test) for steering angle prediction in Autonomous Driving Systems: \textit{Dave\_orig} (DO) and \textit{Dave\_dropout} (DD) 
\cite{Li19}\cite{Pei19}.

\begin{table}[t]
	\centering
	\caption{List of experimental subjects}
	\label{tab:models}
\vspace{-6pt}
\renewcommand{\arraystretch}{.99}
\footnotesize{
\begin{tabular}{c|c|c|c|c}   \toprule
			\textbf{Model} & \textbf{Dataset} & \textbf{Layers} & \textbf{Parameters} & \textbf{Accuracy}\\ \midrule
			A & \multirow{3}{*}{MNIST} & 7 & 6,237 & {90.3\%}\\ \cline{1-1} \cline{3-5} 
			B &  & 6 & 97,114 & {94.8\%} \\ \cline{1-1} \cline{3-5} 
			C &  & 8 & 545,546 & {93.3\%} \\ \hline
			D & \multirow{3}{*}{CIFAR10} & 13 & 1,084,234 & {71.5\%} \\ \cline{1-1} \cline{3-5} 
			E &  & 10 & 258,762 & {79.0\%} \\ \cline{1-1} \cline{3-5} 
			F &  & 12 & 550,570 & {65.1\%} \\ \hline
			G & \multirow{3}{*}{CIFAR100} & 16 & 15,047,588 & {66.3\%} \\ \cline{1-1} \cline{3-5} 
			H &  & 9 & 564,484 & {57.4\%} \\ \cline{1-1} \cline{3-5} 
			I &  & 13 & 1,465,220 & {58.8\%} \\ \hline
            DO & \multirow{2}{*}{Udacity} & 13 & 2,116,983 & 0.904 \\ \cline{1-1} \cline{3-5} 
            DD &  & 15 & 3,276,225 & 0.918 \\ \bottomrule
		\end{tabular}
  }
\vspace{-12pt}
\end{table}

Recht \textit{et al.} \cite{Recht19} showed that if the accuracy is computed on previously unseen data, it is actually smaller than the claimed one by a value ranging from 3\% to 15\% on CIFAR10 and from 11\% to 14\% on ImageNet.
Therefore, for a more realistic accuracy, each DNN is trained ``from scratch'' by separating training, verification, and operational sets, as in \cite{Guerriero23}.
The verification set is the set used to evaluate the DNN. %
The operational set contains unlabelled images. %

The three datasets are split as follows. 
For MNIST, $7,$$000$ images are for training and $2,$$500$ for  verification; the remaining $60,$$500$ entries are the operational dataset (\textit{big size}). All models trained with this configuration achieve an accuracy greater than $90\%$. 
For CIFAR10, we use $24,$$000$ images for training and $2,$$500$ for verification; the remaining $33,$$500$ entries are the operational dataset (\textit{medium size}). 
For CIFAR100, $40,$$000$ entries are for training and $5,000$ for  verification; thus, the operational dataset has $15,$$000$ images (\textit{small size}). 
The operational datasets are chosen to have MNIST (big)  almost double than CIFAR10 (medium) and four times CIFAR100 (small).  
The greater training set sizes for CIFAR10 and CIFAR100 are due to the higher complexity of the images, to pursue an acceptable accuracy. 
For regression models, we use as operational dataset the entire test dataset, as all its examples are unseen during training.

\section{Results}
\label{sect:Results}

\subsection{RQ1: operational accuracy assessment}

\subsubsection{RQ1.1: Classification}
\begin{figure}[t]
    \centering
\includegraphics[width=0.89\columnwidth]{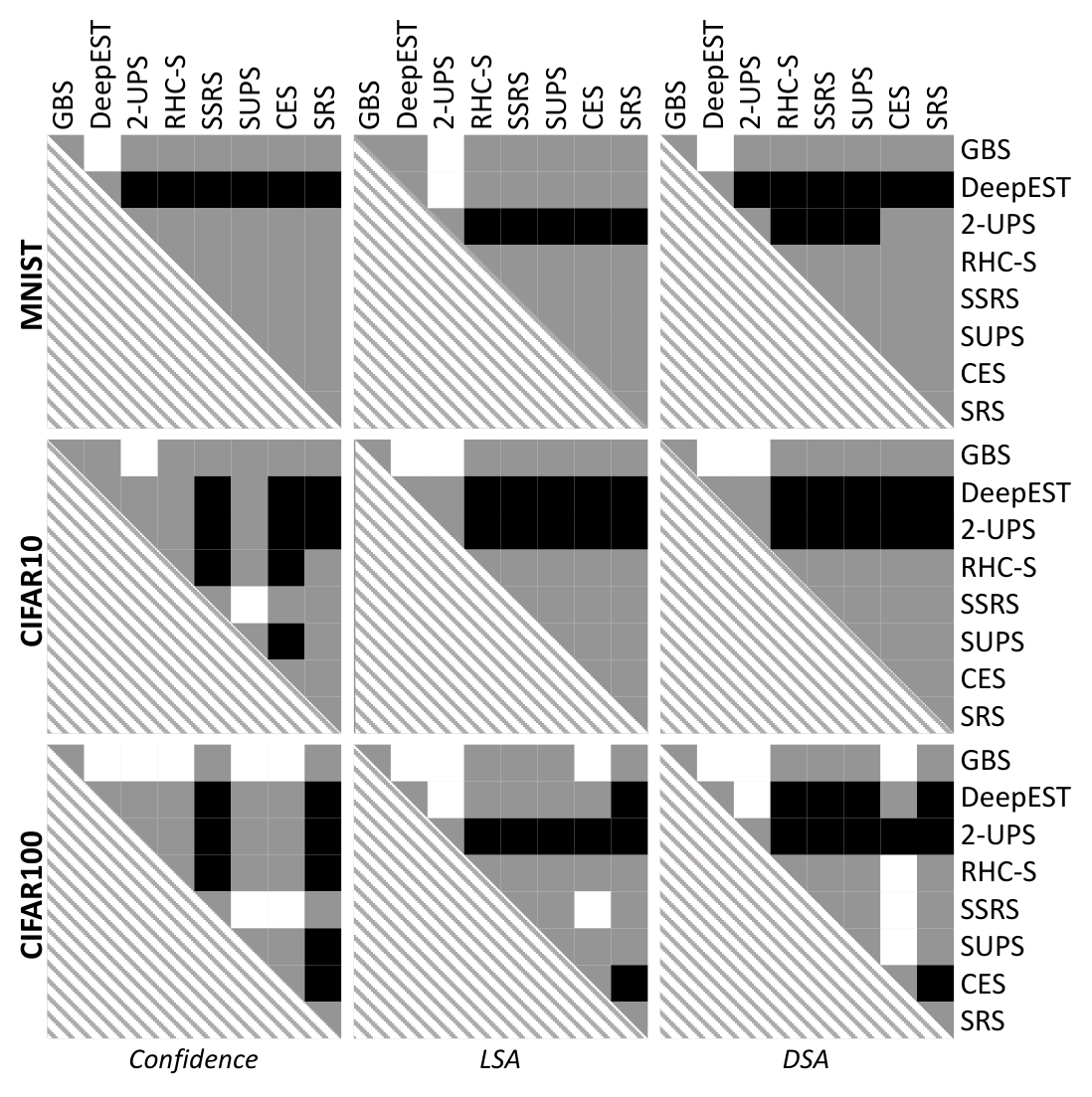}
\vspace{-10pt}
\caption{RQ1.1: Dunn test on the classification task}
    \label{fig:dunn_class}
\vspace{-9pt}
\end{figure}

To check if techniques have pairwise a statistically significant difference, we run the Friedman test \cite{ImanDavenport1980} on all subjects/auxiliary variable pairs. The $p$-value is lower than $\alpha = 0.05$ in all cases, hence the null hypothesis of no difference among techniques is always rejected. 
For pairwise comparison, we run the non-parametric \textit{post hoc} Dunn test \cite{Dinno15} with the Holm adjustment. The results are in Figure \ref{fig:dunn_class}, where gray squares mean \textit{no significant difference} for the pair, white (black) squares mean the technique on the row is statistically better (worse) than the one on the column. All exact $p$-values are in the replication package.\cref{repo}

\begin{figure}[t]
\vspace{-9pt}
	\centering
	\subfloat[Model A with \textit{confidence}]{\label{fig:example1_mnist_conf}
	 	\includegraphics[width=0.23\textwidth]{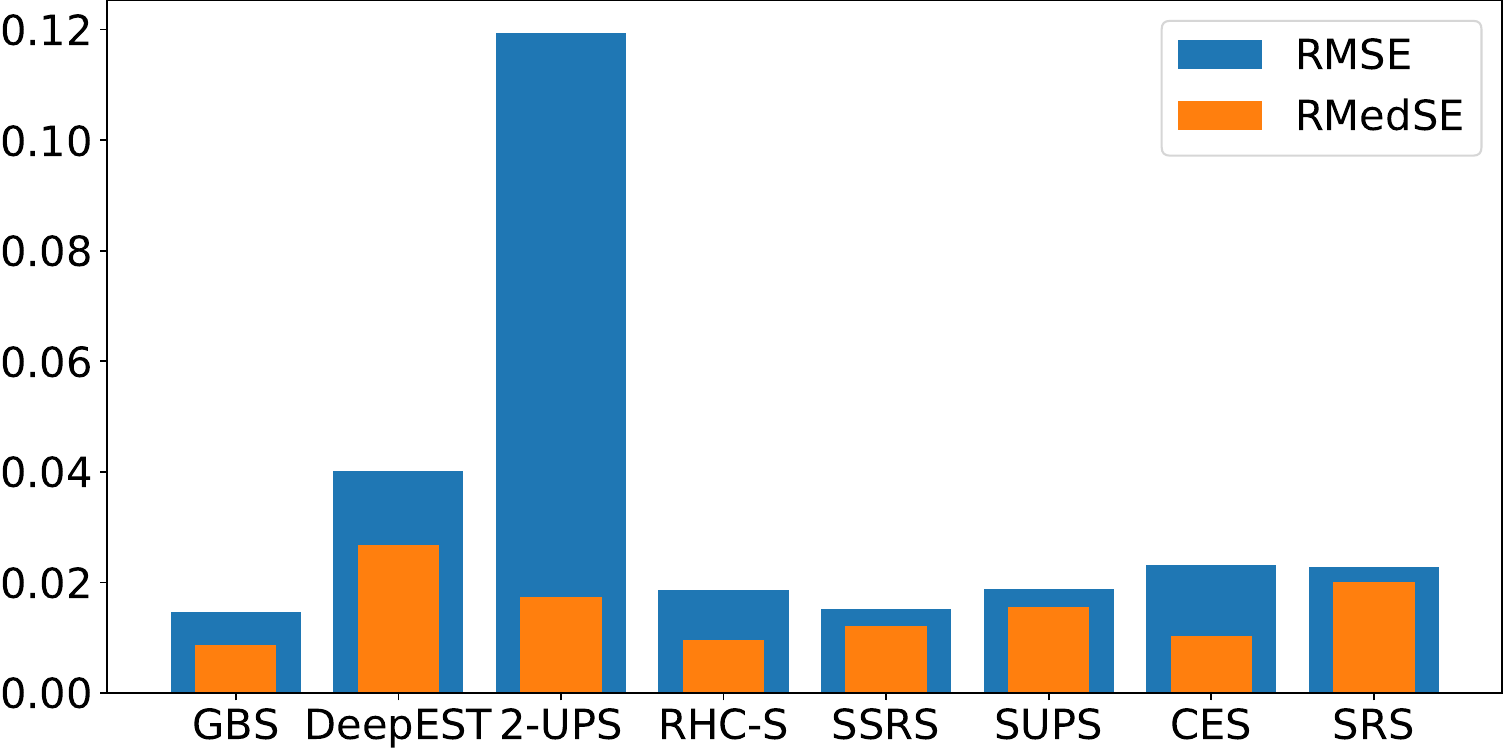}}
   \subfloat[Model B with LSA]{\label{fig:example2_mnist_lsa}
	 	\includegraphics[width=0.23\textwidth]{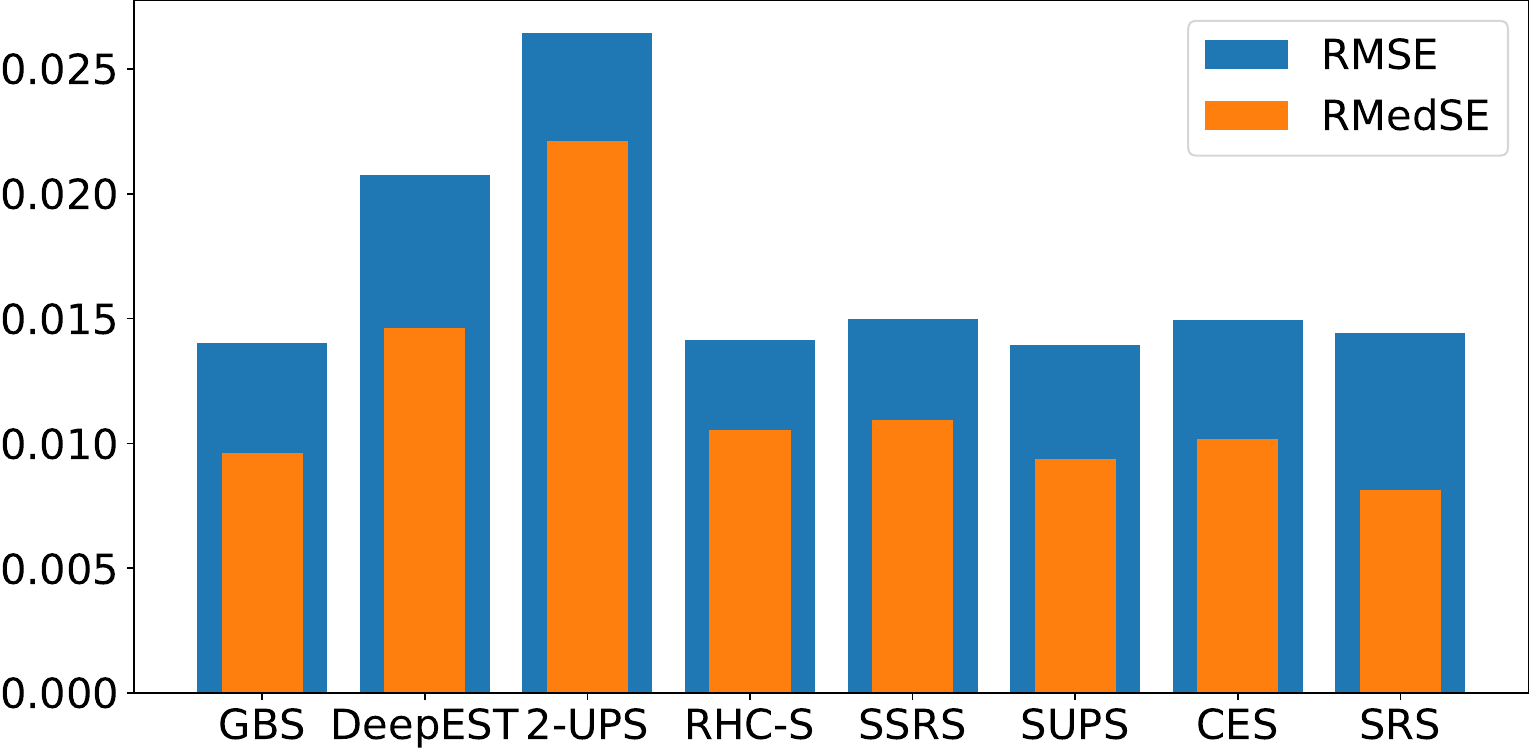}}
   
\vspace{-4pt}
\subfloat[Model C with DSA]{\label{fig:example3_mnist_dsa}
	 	\includegraphics[width=0.23\textwidth]{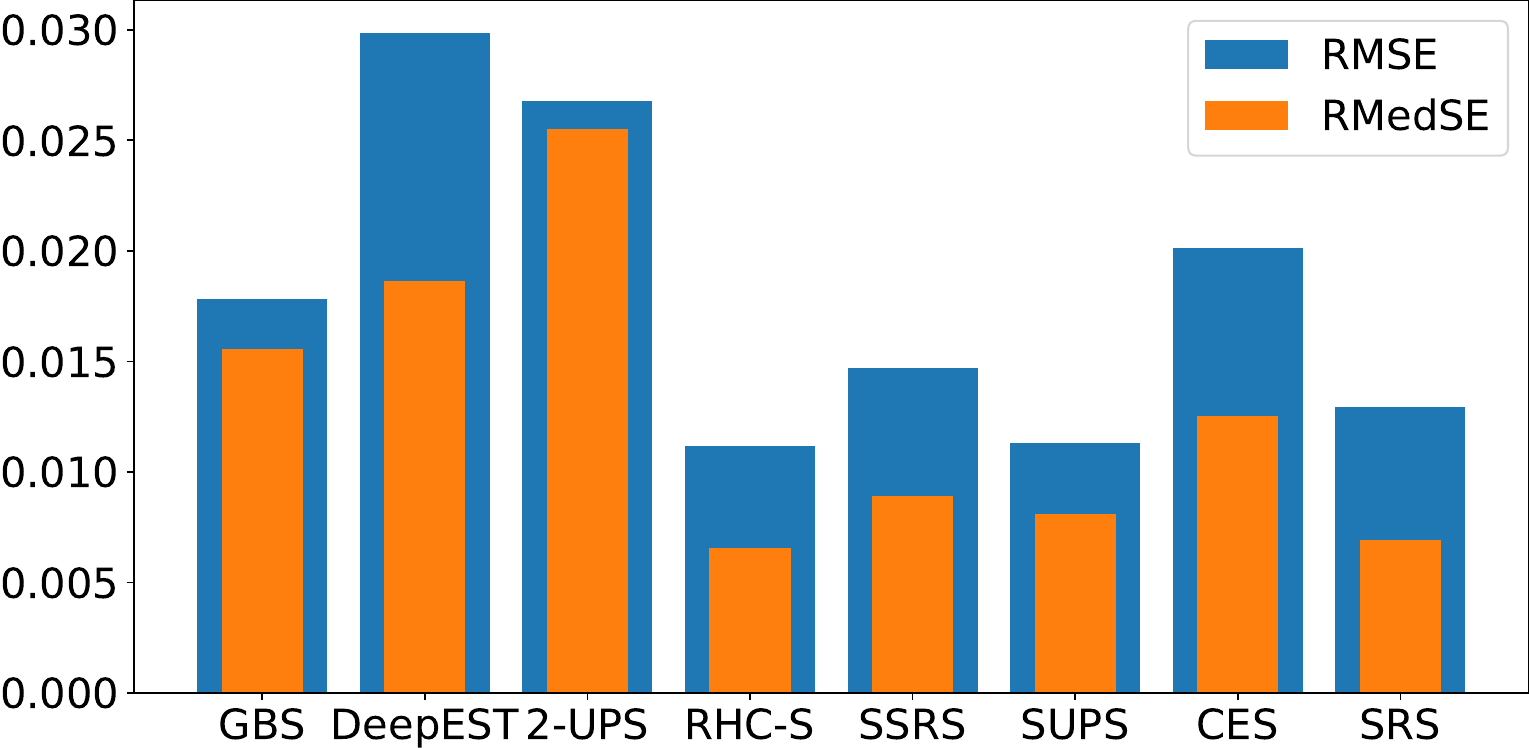}}
   \subfloat[Model E with \textit{confidence}]{\label{fig:example4_cifar_conf}
	 	\includegraphics[width=0.23\textwidth]{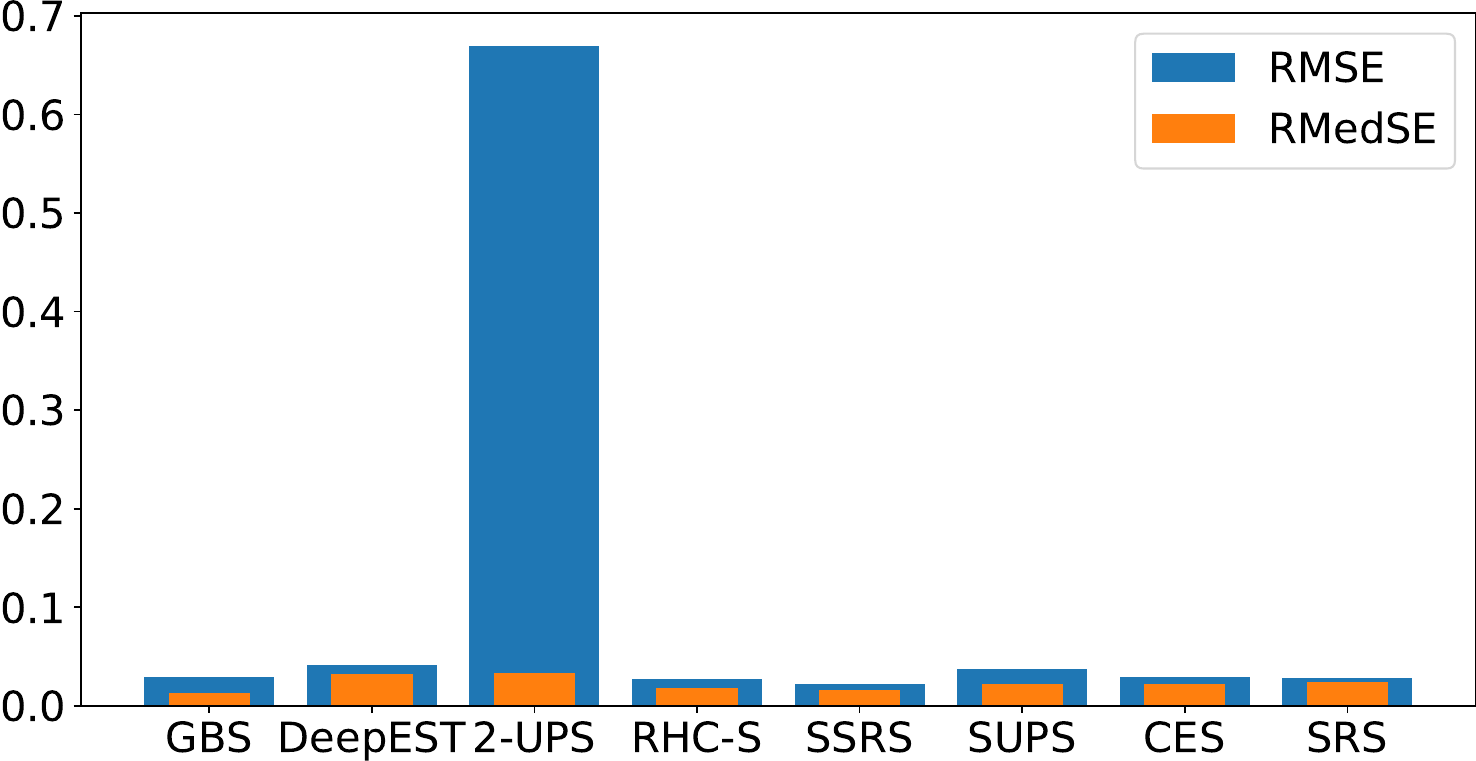}}
   
\vspace{-4pt}
   \subfloat[Model F with DSA]{\label{fig:example5_cifar_dsa}
	 	\includegraphics[width=0.23\textwidth]{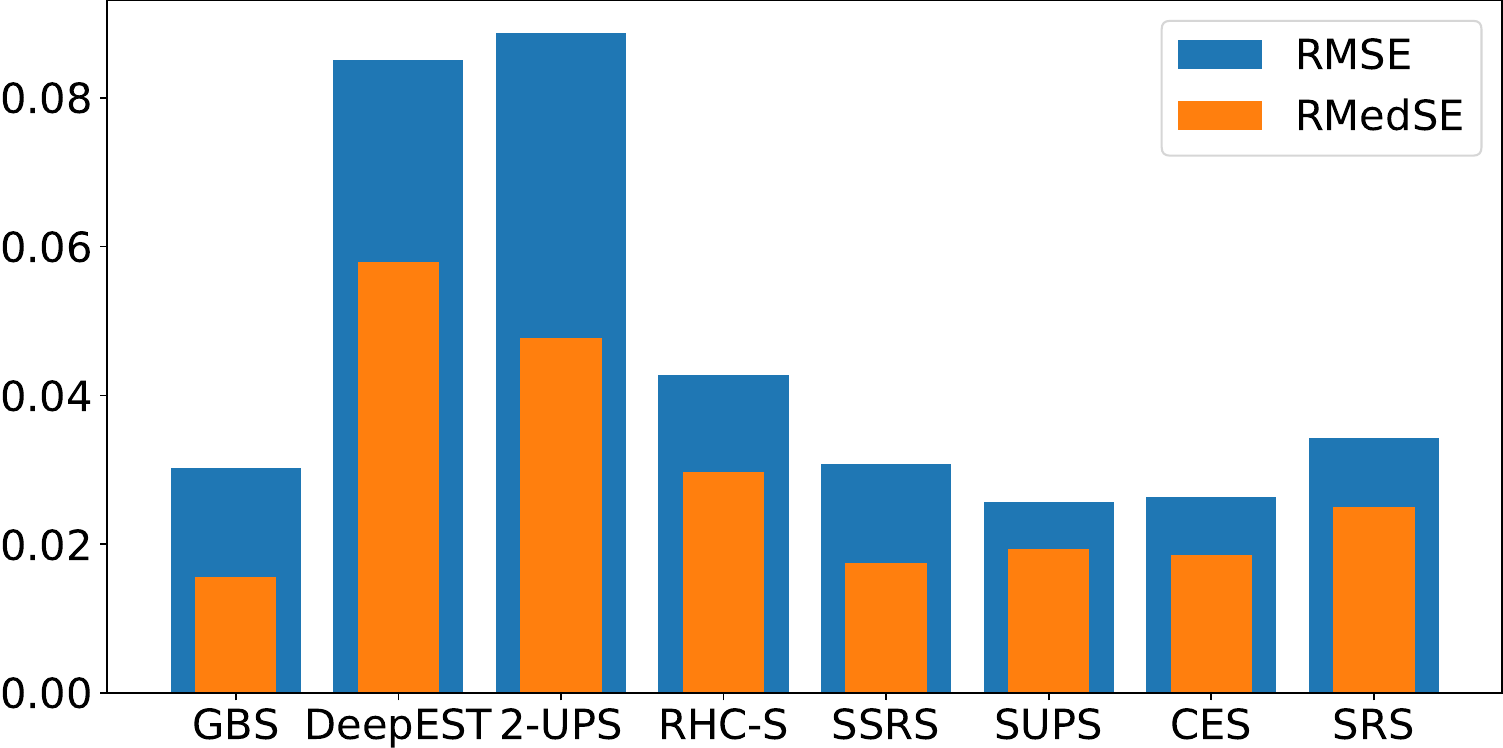}}
   \subfloat[Model I with \textit{confidence}]{\label{fig:example6_cifar100_conf}
	 	\includegraphics[width=0.23\textwidth]{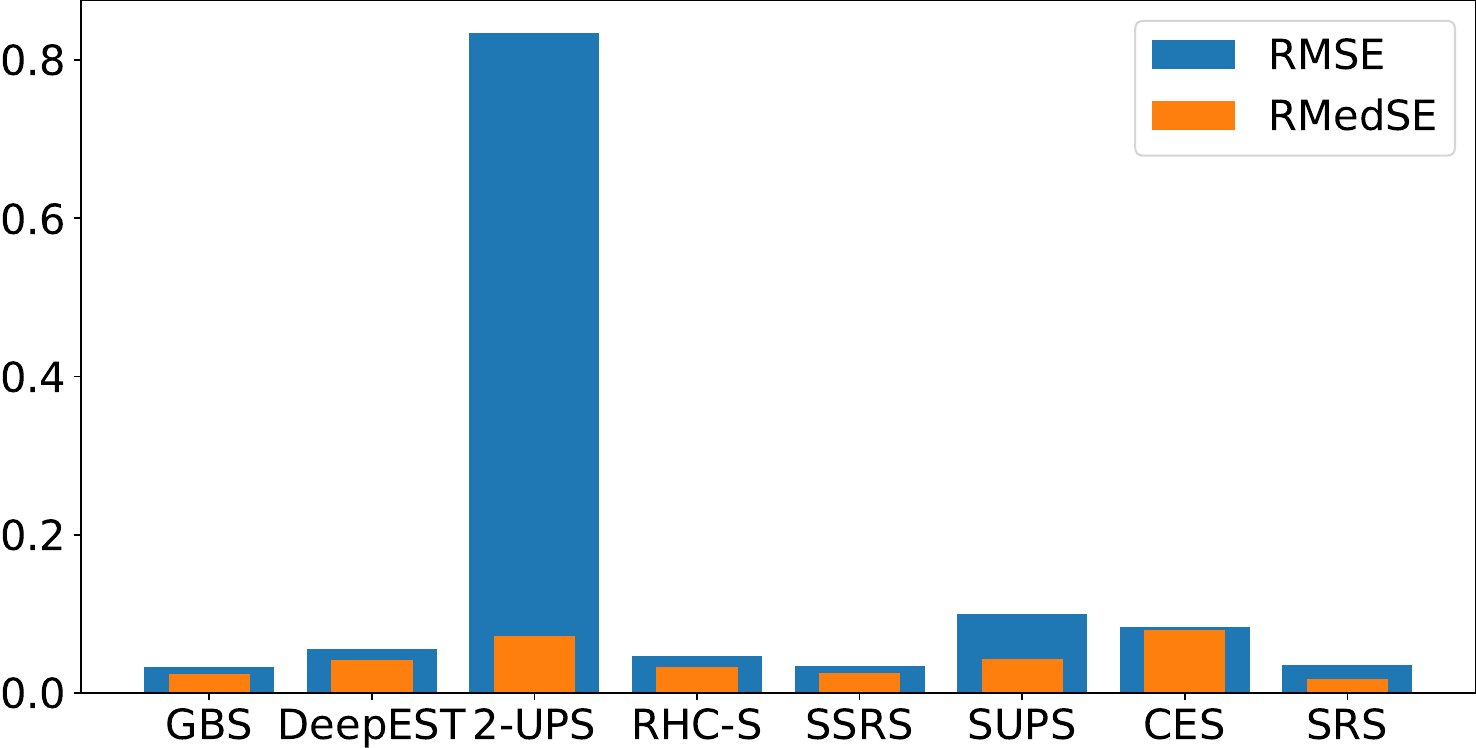}}
   
\vspace{-4pt}
   \subfloat[Model H with LSA]{\label{fig:example7_cifar100_lsa}
	 	\includegraphics[width=0.23\textwidth]{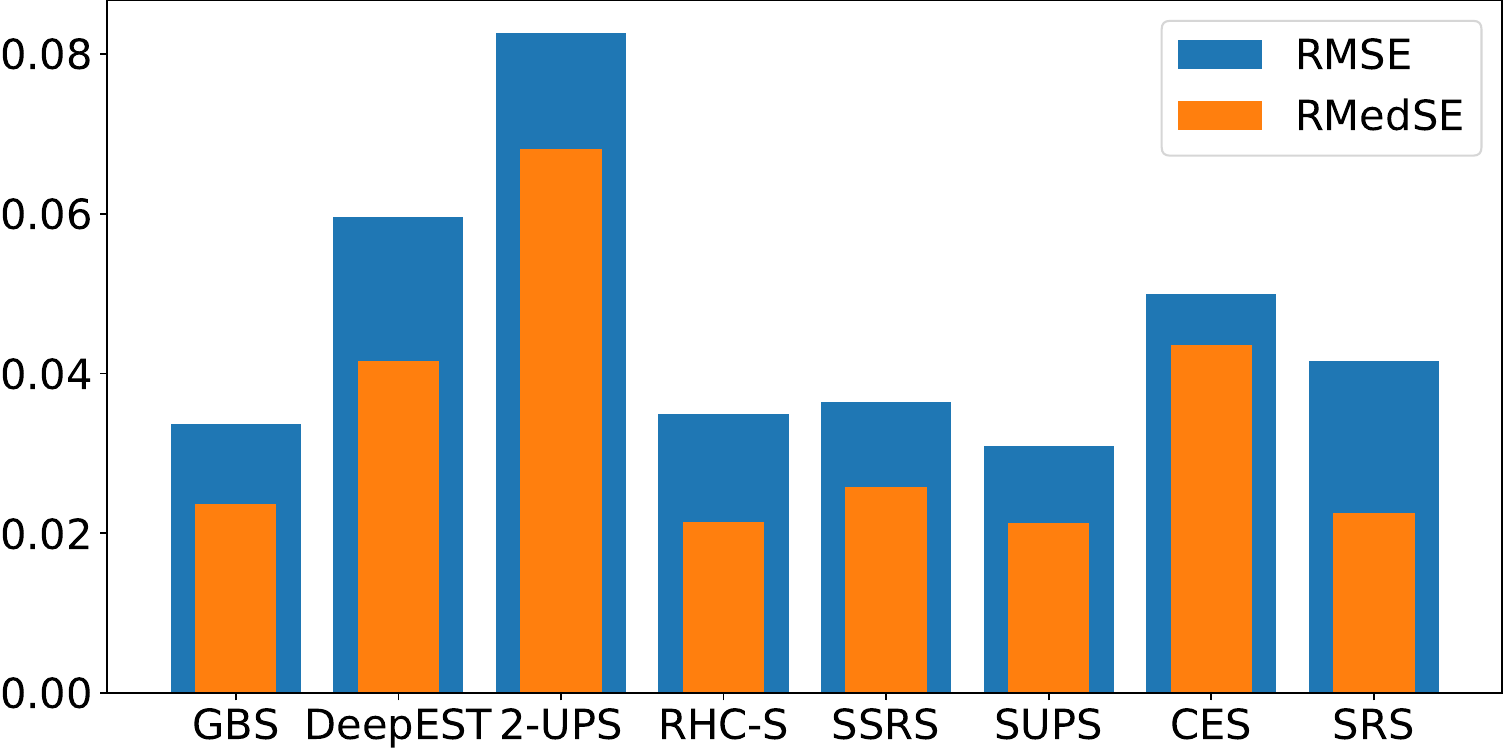}}
   \subfloat[Model H with DSA]{\label{fig:example8_cifar100_dsa}
	 	\includegraphics[width=0.23\textwidth]{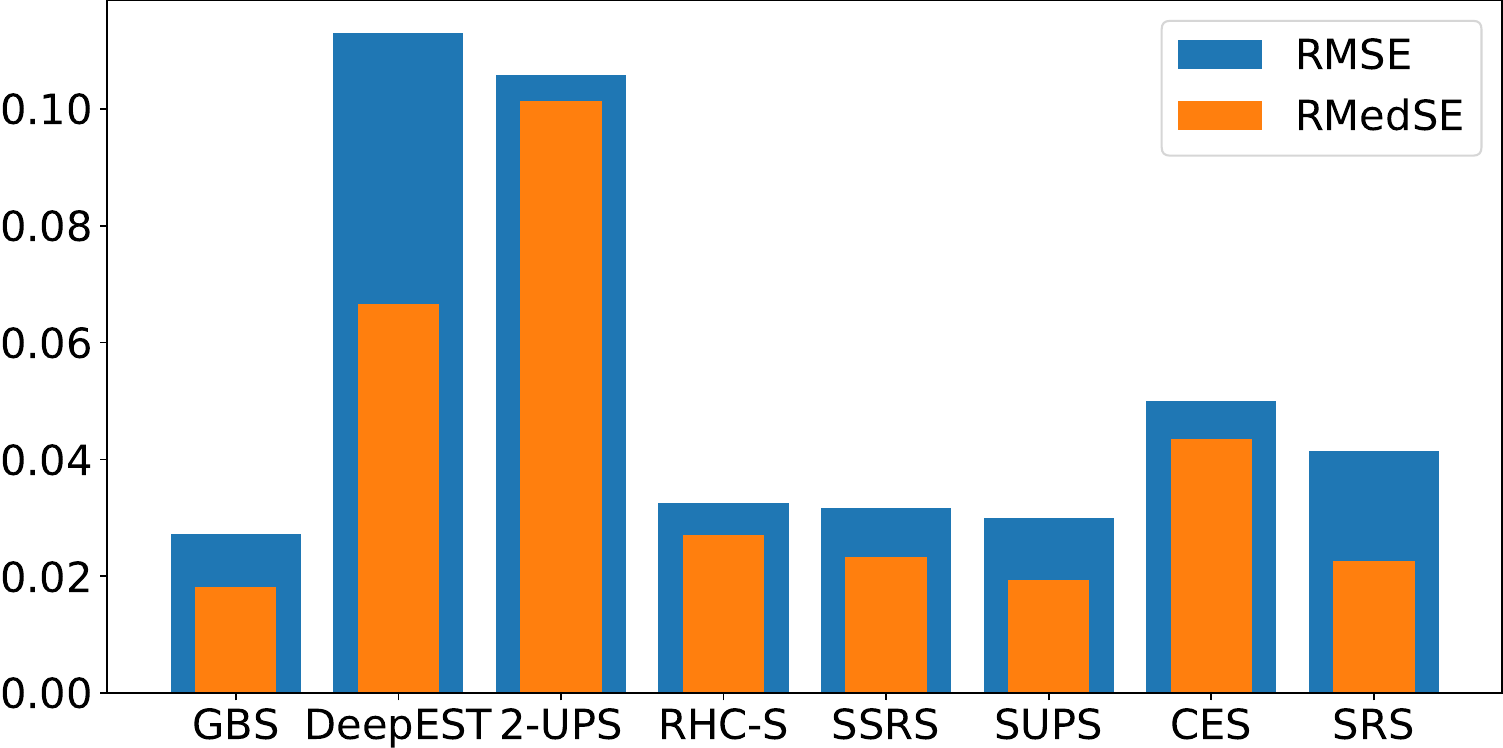}}
\vspace{-9pt}
\caption{RQ1.1: Examples}
	\label{fig:class_bar}
\end{figure}

On MNIST, DeepEST and 2-UPS significantly differ from the other techniques (which perform similarly). We show three examples in Figures \ref{fig:example1_mnist_conf}-\ref{fig:example3_mnist_dsa}. 
The first %
is on Model A (top-left box in Fig. \ref{fig:dunn_class}) with \textit{confidence} as auxiliary variable. Here, the RMSE of 2-UPS is by far the worst; however, it is affected by few outliers due to the inability of the estimator to  balance, within the given budget, the examples whose auxiliary information is incoherent with the result (e.g., failures with high confidence). If we take the root square of the \textit{median} of squared errors, called RMedSE, we see 2-UPS is in line with the others. This causes 2-UPS to go unreported as significantly different by the Dunn test (non-parametric, hence robust to outliers). DeepEST, instead, shows to be significantly worse. 

The second example (Fig. \ref{fig:example2_mnist_lsa}) is on Model B with LSA (top-middle box in Fig. \ref{fig:dunn_class}). 2-UPS performs worse than the others, the second being DeepEST although the difference is not detected by the Dunn test. 
The third example (Fig. \ref{fig:example3_mnist_dsa}) is on Model C with DSA (top-right box in Fig. \ref{fig:dunn_class}). In this case, both DeepEST and 2-UPS perform worse. 
In the second and third examples, the values of the RMSE and RMedSE for 2-UPS are close (no outliers); this is attributable to the higher representativeness of LSA and DSA, which were more robust than \textit{confidence} to misclassification on inputs closer to 
training set.

On CIFAR10 with \textit{confidence}, the outliers in 2-UPS are even more pronounced (Fig. \ref{fig:example4_cifar_conf}). 
DeepEST and 2-UPS again give the worst estimates. %
The other algorithms are similar (Fig. \ref{fig:example5_cifar_dsa}). %

On CIFAR100 with \textit{confidence}, GBS, SSRS and SRS differ significantly from the other techniques. Consider Fig. \ref{fig:example6_cifar100_conf} (Model I). GBS, SSRS and SRS exhibit the best values. Outliers in 2-UPS are confirmed; they are more frequent, especially 
on low-accurate models (the behaviour is more evident with CIFAR10 and CIFAR100, less accurate for MNIST).
On the other hand, it is worth to stress that not all the algorithms relying on the auxiliary variable suffer from unstable results; RHC-S and SUPS are more stable. 
With LSA (Fig. \ref{fig:example7_cifar100_lsa}) and DSA (Fig. \ref{fig:example8_cifar100_dsa}), the previous results are confirmed; after DeepEST and 2-UPS, CES turned out to be the third worst one.

\subsubsection{RQ1.2: Regression}
\begin{figure}[t]
    \centering
\includegraphics[width=0.89\columnwidth]{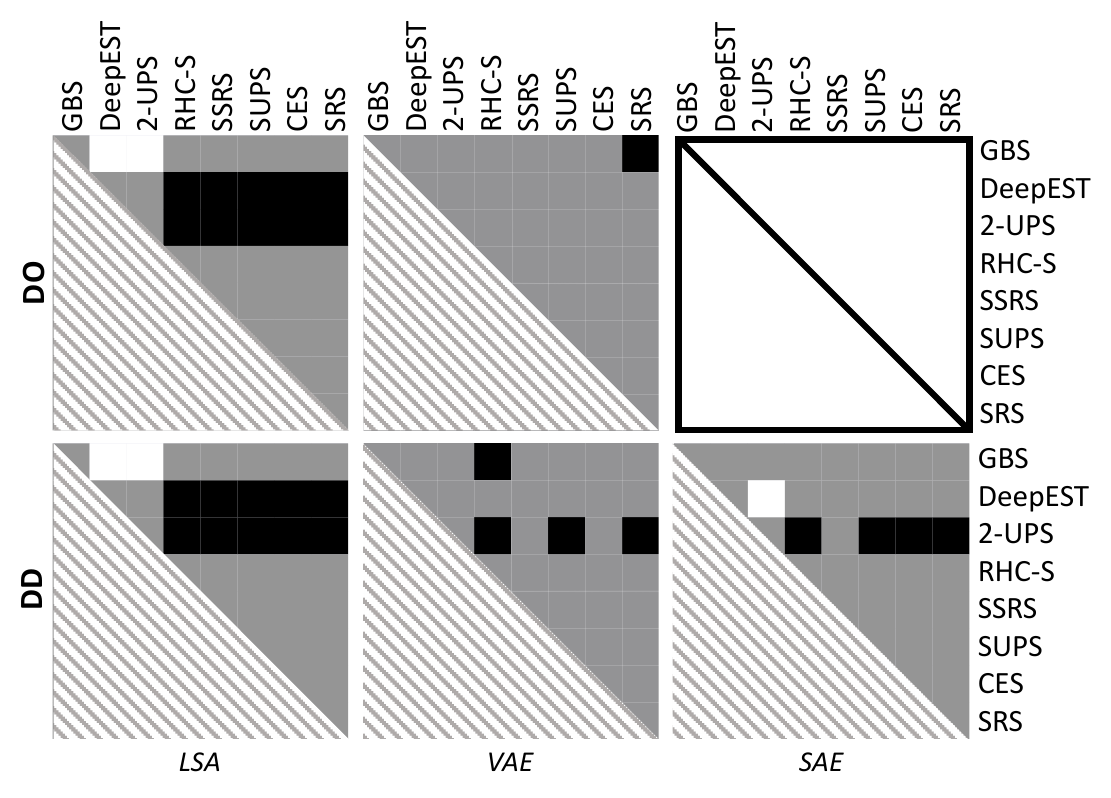}
\vspace{-9pt}
    \caption{RQ1.2: Dunn test on the regression task}
    \label{fig:dunn_reg}
\vspace{-10pt}
\end{figure}

\begin{figure}[t]
\vspace{-9pt}
	\centering
	\subfloat[LSA]{
	 	\includegraphics[width=0.23\textwidth]{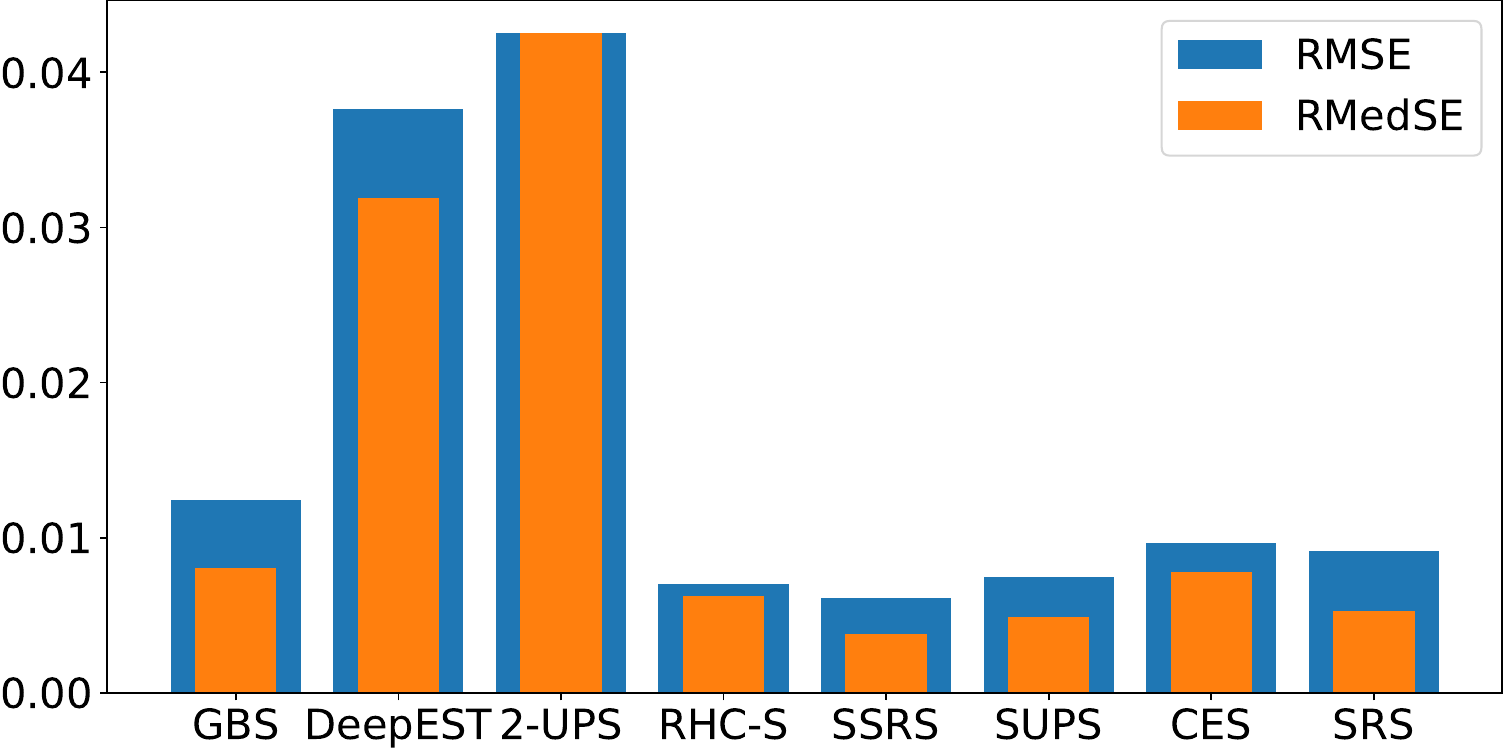}}
   \subfloat[VAE]{
	 	\includegraphics[width=0.23\textwidth]{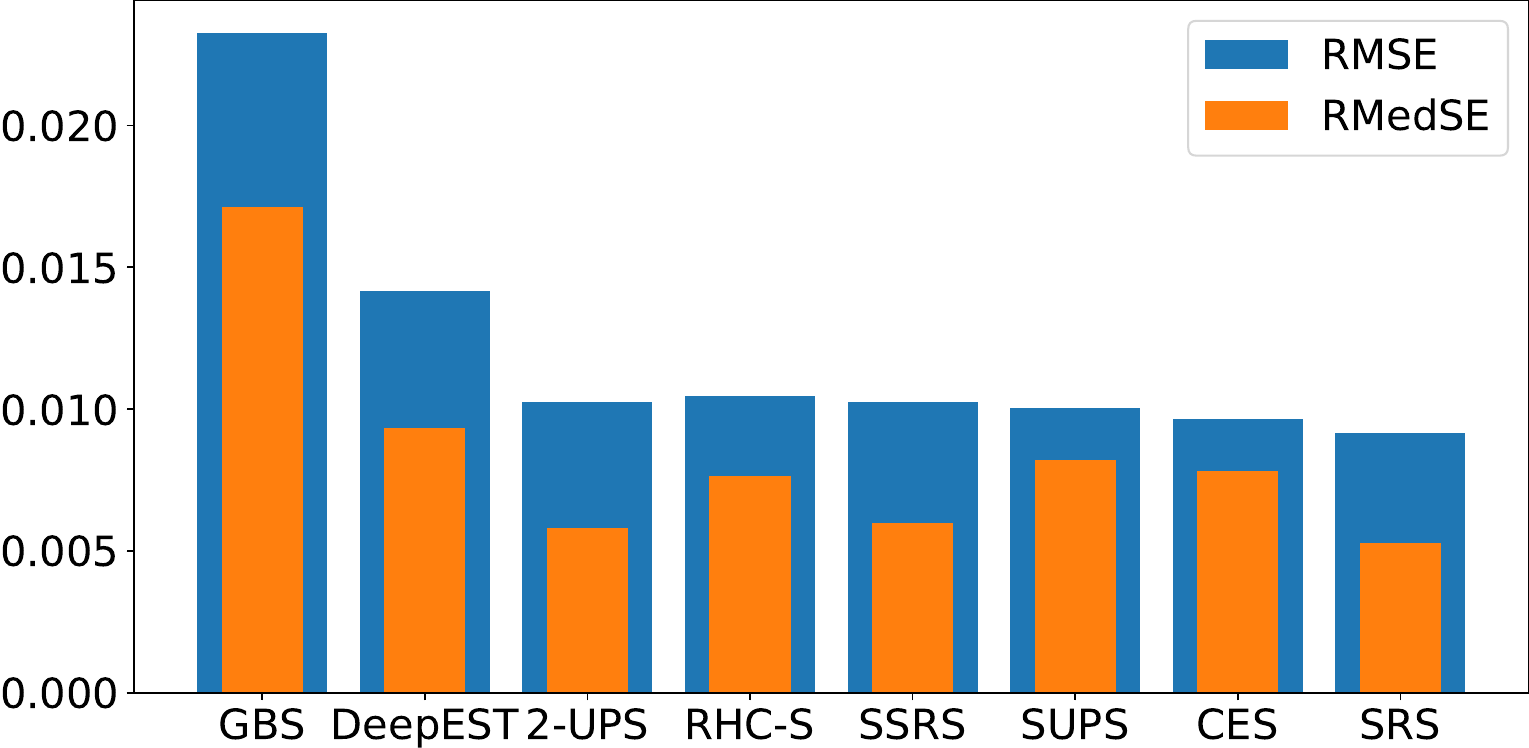}}
 \vspace{-6pt}
  
   \subfloat[SAE]{
	 	\includegraphics[width=0.23\textwidth]{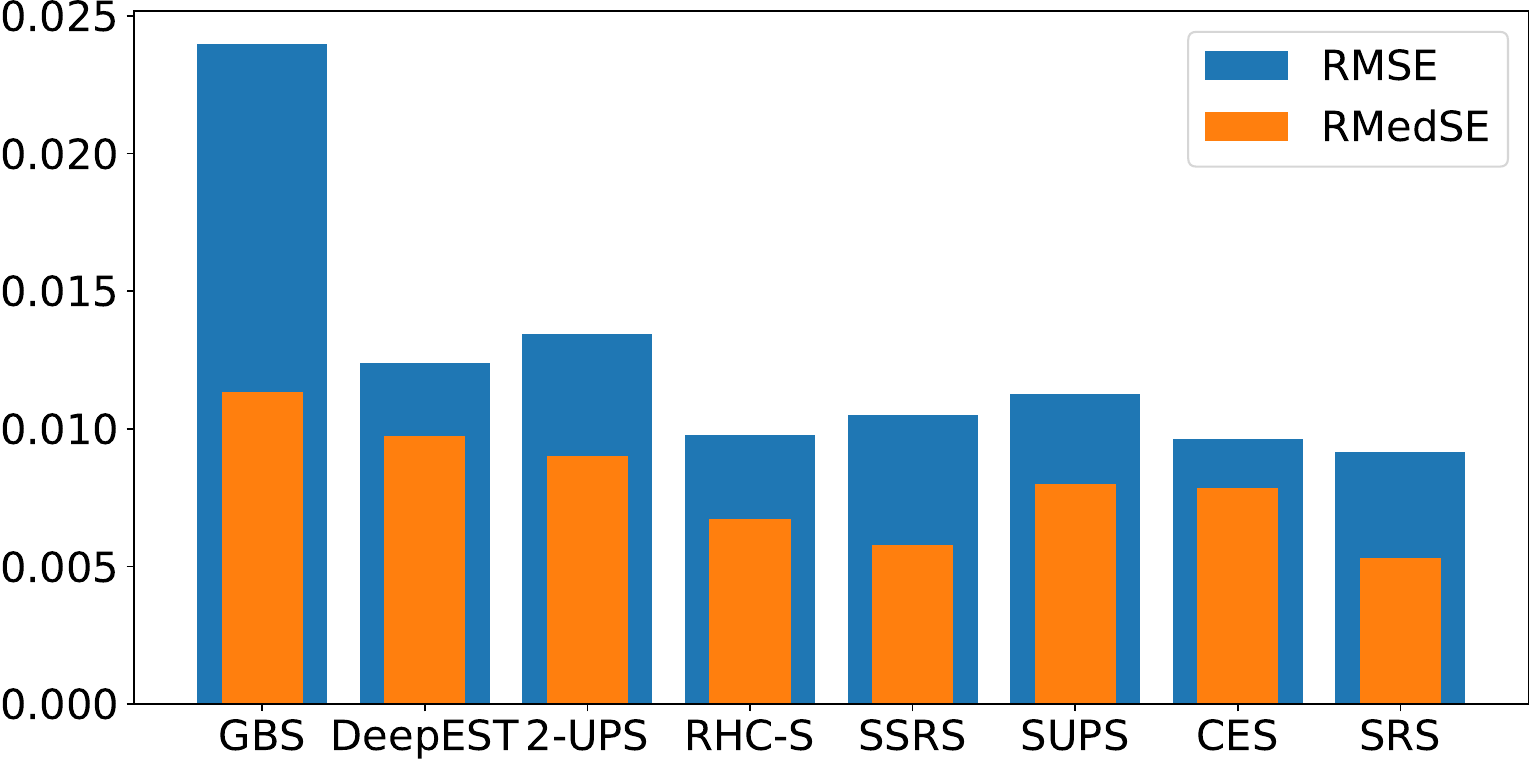}}
\vspace{-9pt}
\caption{RQ1.2: DO model - Bar charts}
	\label{fig:reg_bar_do}
\end{figure}

\begin{figure}[t]
	\centering
	\subfloat[LSA]{
	 	\includegraphics[width=0.23\textwidth]{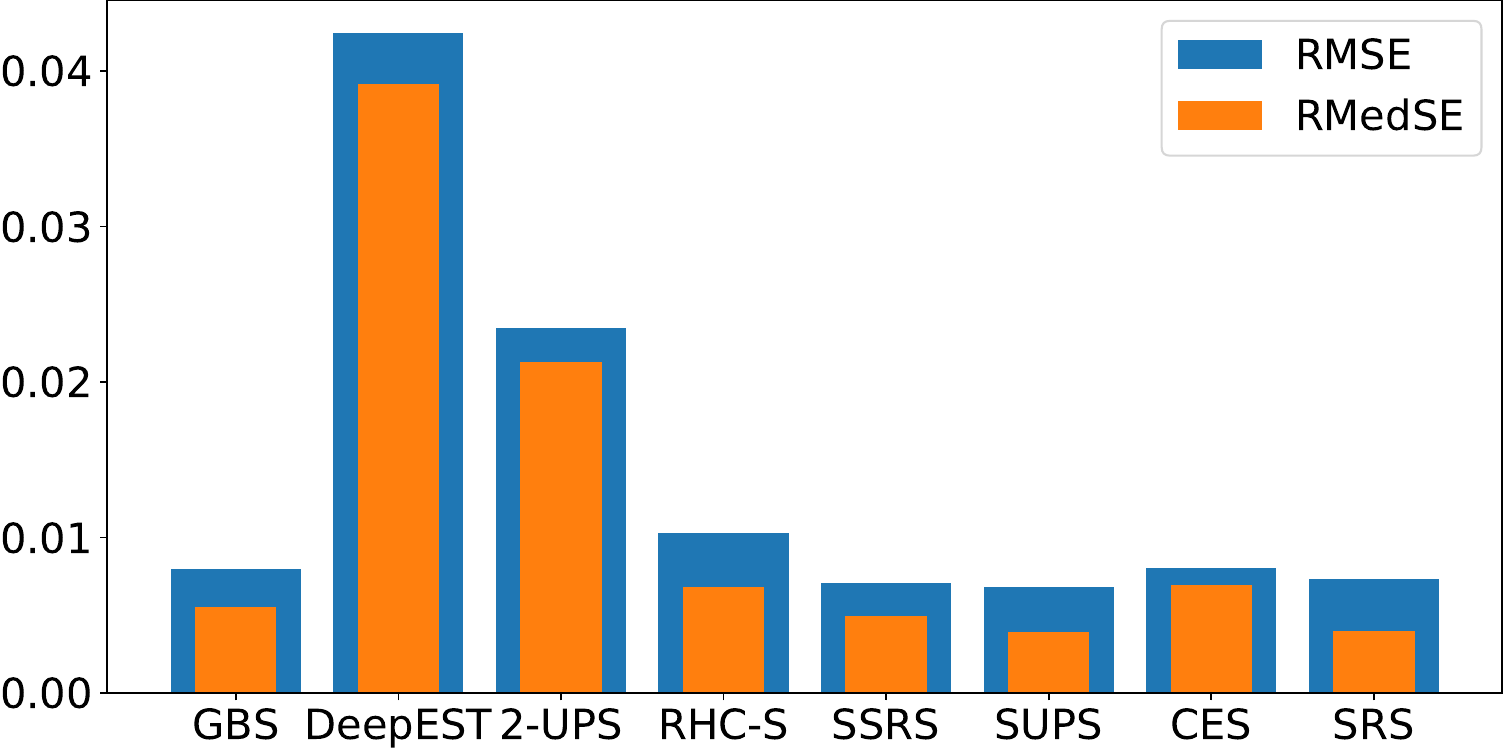}}
   \subfloat[VAE]{
	 	\includegraphics[width=0.23\textwidth]{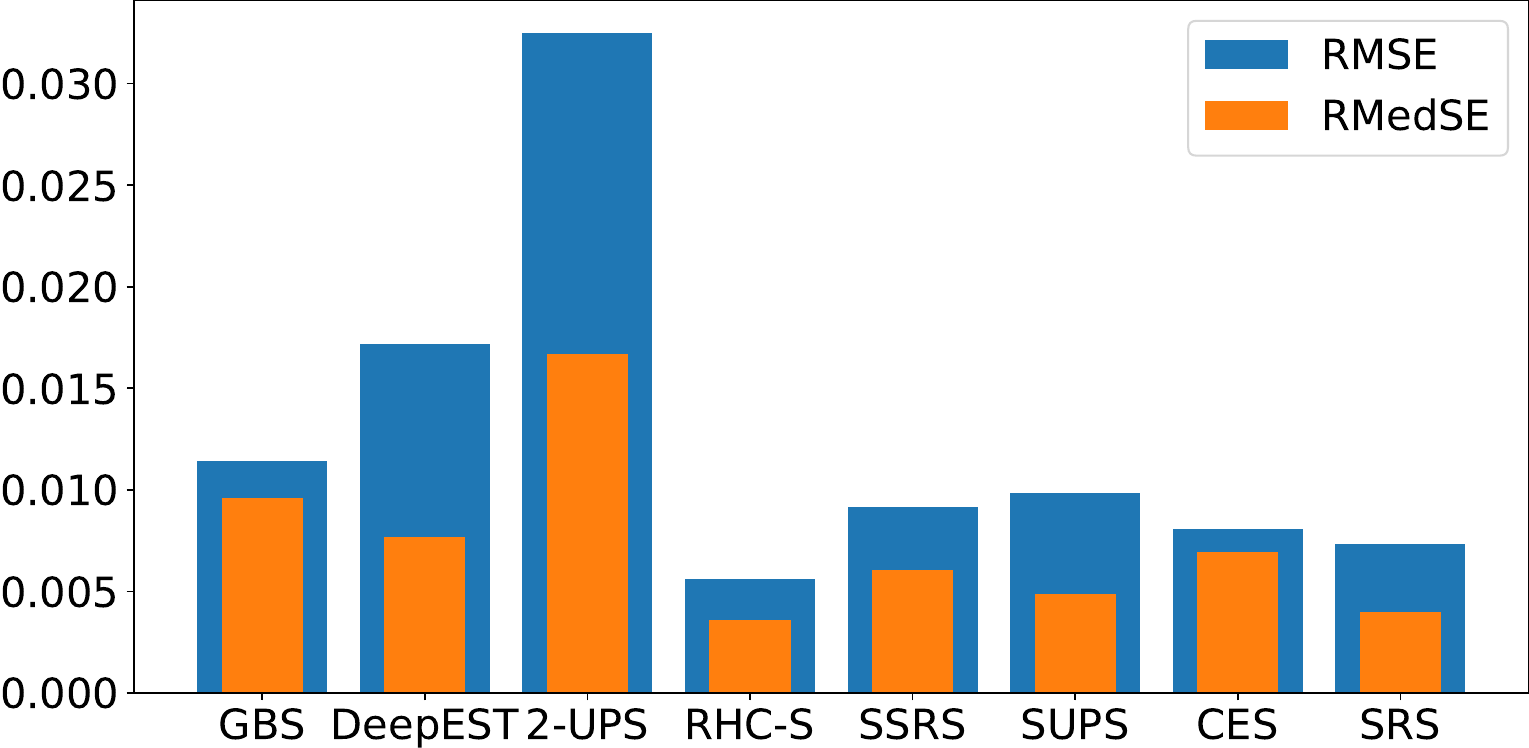}}
   \vspace{-6pt}

   \subfloat[SAE]{
	 	\includegraphics[width=0.23\textwidth]{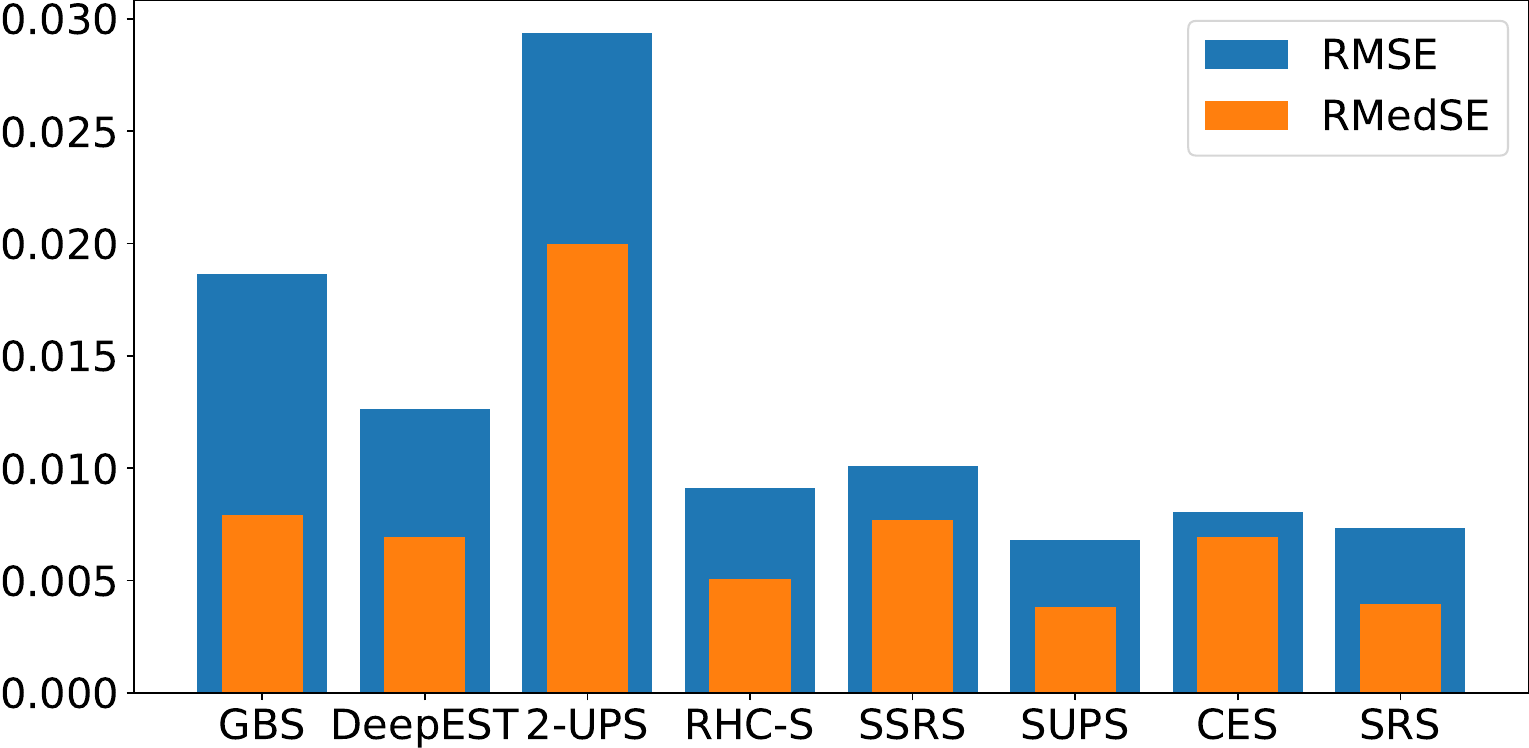}}
\vspace{-9pt}
\caption{RQ1.2: DD model - Bar charts}
	\label{fig:reg_bar_dd}
\vspace{-12pt}
\end{figure}

The Friedman test gives a $p$-value lower than $\alpha = 0.05$ in all the cases, except for DO with the SAE auxiliary variable. %
Figure \ref{fig:dunn_reg} shows the results of the Dunn test for pairwise comparison. 
When using LSA, DeepEST and 2-UPS are significantly worse. Figures \ref{fig:reg_bar_do} and \ref{fig:reg_bar_dd} %
confirm their higher values of RMSE and RMedSE for both DO and DD models. %

With the VAE auxiliary variable, GBS differs from SRS, but it is almost equivalent to the other algorithms in the DO model. 
Figure \ref{fig:reg_bar_do} confirms that GBS has  higher RMSE than the others. For the DD model, 2-UPS differs from SUPS and RHC-S. 
2-UPS has the highest RMSE values, while RHC-S has the lowest ones (Figure \ref{fig:reg_bar_dd}). 

With SAE, the Friedman test did not detect any difference for DO, while, for DD, 2-UPS is still the worst technique, although it is closer to GBS and SSRS than the previous cases. 

As for RMedSE\%, the worst values are always with the LSA variable: for DO, 2-UPS and DeepEST show the worst values (\textbf{4.2$\%$} and \textbf{3.2$\%$}, respectively); SSRS has the best value (\textbf{0.4$\%$}); for DD, DeepEST and 2-UPS show \textbf{3.9$\%$} and \textbf{2.1$\%$}, respectively, against SUPS with \textbf{0.4$\%$}. The worst case with autoencoders is for DD with the SAE variable, with 2-UPS (\textbf{2.0$\%$}), while the best one is SRS (\textbf{0.4$\%$}). These results are in line with the classification ones.

Overall, the techniques are all equivalently effective in assessing the operational accuracy of DNN models for classification and regression tasks, except for DeepEST and 2-UPS. While DeepEST was expected to show worse results (its primary objective is on failure exposure), 2-UPS shows many outliers since it is strongly affected by auxiliary variable representativeness.

\subsection{RQ2: failing examples detection}
For RQ2, we treat  classification and regression  differently. For the former, we count the number of misclassifications. For the latter, we count the number of examples whose offset $\delta$ (predicted \textit{vs} actual value) is greater than a threshold $y$, with $y \in [0^\circ, 2.5^\circ, 5^\circ, \dots, 25^\circ]$ -- the higher the difference, the more severe the misprediction. %

\subsubsection{RQ2.1: Classification}
Table \ref{tab:fails_class} reports the number of misclassifications broken down by dataset and auxiliary variable -- the best mean values are in bold. DeepEST exposes more failures than the others in 7 out of 9 times.  2-UPS has the highest value only for CIFAR10 with confidence, and %
SUPS for CIFAR100 with confidence. 2-UPS, SUPS, and RHC-S almost equivalently follow DeepEST. 

\begin{table}[t]
\centering
\caption{RQ2.1: Number of exposed failures (classification)}%
\label{tab:fails_class}
\setlength{\tabcolsep}{7pt}
\vspace{-6pt}
\renewcommand{\arraystretch}{1.}
\small{
\resizebox{\columnwidth}{!}{
\begin{tabular}{cc||cc||cc||cc}   \toprule
 & {} & \multicolumn{2}{c||}{\textbf{MNIST}} & \multicolumn{2}{c||}{\textbf{CIFAR10}} & \multicolumn{2}{c}{\textbf{CIFAR100}} \\ \hline 
\multicolumn{1}{c|}{\textit{$\chi$}} & Technique & \multicolumn{1}{c|}{\textit{mean}} & \textit{std} & \multicolumn{1}{c|}{\textit{mean}} & \textit{std} & \multicolumn{1}{c|}{\textit{mean}} & \textit{std} \\ \hline\hline
\multicolumn{1}{c|}{\multirow{8}{*}{\rotatebox[origin=c]{90}{\textbf{confidence}}}} & GBS & \multicolumn{1}{c|}{28.2} & 7.2 & \multicolumn{1}{c|}{68.4} & 8.0 & \multicolumn{1}{c|}{86.2} & 10.6 \\ \cline{2-8} 
\multicolumn{1}{c|}{} & DeepEST & \multicolumn{1}{c|}{\textbf{80.5}} & 10.3 & \multicolumn{1}{c|}{108.7} & 9.4 & \multicolumn{1}{c|}{136.4} & 6.6 \\ \cline{2-8} 
\multicolumn{1}{c|}{} & 2-UPS & \multicolumn{1}{c|}{69.5} & 17.4 & \multicolumn{1}{c|}{\textbf{108.9}} & 11.7 & \multicolumn{1}{c|}{141.5} & 6.2 \\ \cline{2-8} 
\multicolumn{1}{c|}{} & RHC-S & \multicolumn{1}{c|}{70.6} & 16.5 & \multicolumn{1}{c|}{106.0} & 12.8 & \multicolumn{1}{c|}{{142.3}} & 6.4 \\ \cline{2-8} 
\multicolumn{1}{c|}{} & SSRS & \multicolumn{1}{c|}{38.4} & 10.2 & \multicolumn{1}{c|}{78.7} & 13.5 & \multicolumn{1}{c|}{109.2} & 6.2 \\ \cline{2-8}
\multicolumn{1}{c|}{} & SUPS & \multicolumn{1}{c|}{69.8} & 16.9 & \multicolumn{1}{c|}{106.9} & 12.3 & \multicolumn{1}{c|}{\textbf{143.6}} & 5.5 \\ \cline{2-8} 
\multicolumn{1}{c|}{} & CES & \multicolumn{1}{c|}{15.6} & 5.8 & \multicolumn{1}{c|}{55.8} & 12.6 & \multicolumn{1}{c|}{70.4} & 7.5 \\ \cline{2-8} 
\multicolumn{1}{c|}{} & SRS & \multicolumn{1}{c|}{14.6} & 5.3 & \multicolumn{1}{c|}{57.3} & 13.1 & \multicolumn{1}{c|}{78.0} & 12.8 \\ \hline\hline
\multicolumn{1}{c|}{\multirow{8}{*}{\rotatebox[origin=c]{90}{\textbf{LSA}}}} & GBS & \multicolumn{1}{c|}{21.0} & 5.2 & \multicolumn{1}{c|}{58.2} & 10.6 & \multicolumn{1}{c|}{84.6} & 7.4 \\ \cline{2-8} 
\multicolumn{1}{c|}{} & DeepEST & \multicolumn{1}{c|}{\textbf{35.9}} & 10.7 & \multicolumn{1}{c|}{\textbf{69.2}} & 10.4 & \multicolumn{1}{c|}{\textbf{119.9}} & 12.4 \\ \cline{2-8} 
\multicolumn{1}{c|}{} & 2-UPS & \multicolumn{1}{c|}{25.0} & 7.5 & \multicolumn{1}{c|}{61.8} & 11.3 & \multicolumn{1}{c|}{110.5} & 20.1 \\ \cline{2-8} 
\multicolumn{1}{c|}{} & RHC-S & \multicolumn{1}{c|}{25.5} & 7.1 & \multicolumn{1}{c|}{62.9} & 11.4 & \multicolumn{1}{c|}{110.3} & 19.3 \\ \cline{2-8} 
\multicolumn{1}{c|}{} & SSRS & \multicolumn{1}{c|}{27.3} & 6.4 & \multicolumn{1}{c|}{60.1} & 10.7 & \multicolumn{1}{c|}{93.2} & 10.7 \\ \cline{2-8} 
\multicolumn{1}{c|}{} & SUPS & \multicolumn{1}{c|}{25.4} & 6.8 & \multicolumn{1}{c|}{63.5} & 11.2 & \multicolumn{1}{c|}{110.2}  & 21.7 \\ \cline{2-8} 
\multicolumn{1}{c|}{} & CES & \multicolumn{1}{c|}{15.6} & 5.8 & \multicolumn{1}{c|}{55.8} & 12.6 & \multicolumn{1}{c|}{70.4} & 7.5 \\ \cline{2-8} 
\multicolumn{1}{c|}{} & SRS & \multicolumn{1}{c|}{14.6} & 5.3 & \multicolumn{1}{c|}{57.3} & 13.1 & \multicolumn{1}{c|}{78.0}  & 12.8 \\ \hline\hline
\multicolumn{1}{c|}{\multirow{8}{*}{\rotatebox[origin=c]{90}{\textbf{DSA}}}} & GBS & \multicolumn{1}{c|}{20.3} & 6.0 & \multicolumn{1}{c|}{63.6} & 11.0 & \multicolumn{1}{c|}{88.5} & 7.5 \\ \cline{2-8} 
\multicolumn{1}{c|}{} & DeepEST & \multicolumn{1}{c|}{\textbf{73.0}} & 17.9 & \multicolumn{1}{c|}{\textbf{102.4}} & 8.9 & \multicolumn{1}{c|}{\textbf{136.7}} & 4.9 \\ \cline{2-8} 
\multicolumn{1}{c|}{} & 2-UPS & \multicolumn{1}{c|}{25.2} & 7.2 & \multicolumn{1}{c|}{65.2} & 12.0 & \multicolumn{1}{c|}{96.3} & 8.7 \\ \cline{2-8} 
\multicolumn{1}{c|}{} & RHC-S & \multicolumn{1}{c|}{23.9} & 7.0 & \multicolumn{1}{c|}{65.5} & 13.8 & \multicolumn{1}{c|}{96.9} & 9.6 \\ \cline{2-8} 
\multicolumn{1}{c|}{} & SSRS & \multicolumn{1}{c|}{21.7} & 5.6 & \multicolumn{1}{c|}{58.3} & 12.0 & \multicolumn{1}{c|}{82.4} & 6.4 \\ \cline{2-8} 
\multicolumn{1}{c|}{} & SUPS & \multicolumn{1}{c|}{24.7} & 6.9 & \multicolumn{1}{c|}{65.7} & 12.1 & \multicolumn{1}{c|}{97.3} & 10.7 \\ \cline{2-8} 
\multicolumn{1}{c|}{} & CES & \multicolumn{1}{c|}{15.6} & 5.8 & \multicolumn{1}{c|}{55.8} & 12.6 & \multicolumn{1}{c|}{70.4} & 7.5 \\ \cline{2-8} 
\multicolumn{1}{c|}{} & SRS & \multicolumn{1}{c|}{14.6} & 5.3 & \multicolumn{1}{c|}{57.3} & 13.1 & \multicolumn{1}{c|}{78.0} & 12.8 \\    \bottomrule
\end{tabular}}}
\vspace{-9pt}
\end{table}

\begin{table*}[]
\centering
\caption{RQ2.2: Average number of failures whose offset is
$ 12.5^\circ \leq \delta< 15^\circ$, $ 15^\circ \leq \delta< 17.5^\circ$, \dots $ 22.5^\circ \leq \delta< 25^\circ$. DO model}
\label{tab:fails_reg_DO}
\vspace{-6pt}
\footnotesize{
\resizebox{0.9\textwidth}{!}{
\begin{tabular}{c|cc|cc|c|c} \toprule
{Technique} &  \multicolumn{2}{c|}{LSA} & \multicolumn{2}{c|}{VAE} & \multicolumn{2}{c}{SAE}\\ \hline
\renewcommand{\arraystretch}{1.4}
GBS & \multicolumn{1}{c|}{\includegraphics[width=0.20\textwidth]{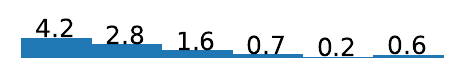}} & 10.2 & \multicolumn{1}{c|}{\includegraphics[width=0.20\textwidth]{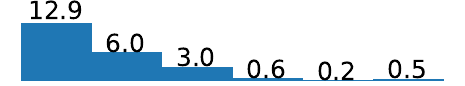}} & \textbf{23.1} & \includegraphics[width=0.20\textwidth]{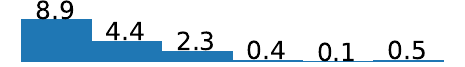} & 16.6 \\ \hline
DeepEST & \multicolumn{1}{c|}{\includegraphics[width=0.20\textwidth]{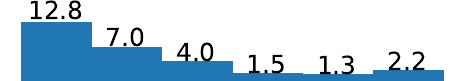}} & 28.9 & \multicolumn{1}{c|}{\includegraphics[width=0.20\textwidth]{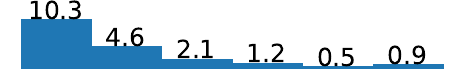}} & 19.5 & \includegraphics[width=0.20\textwidth]{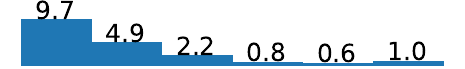} & \textbf{19.2} \\ \hline
2-UPS & \multicolumn{1}{c|}{\includegraphics[width=0.20\textwidth]{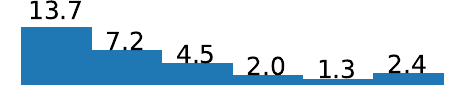}} & 31.0 & \multicolumn{1}{c|}{\includegraphics[width=0.20\textwidth]{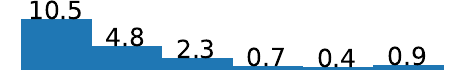}} & 19.6 & \includegraphics[width=0.20\textwidth]{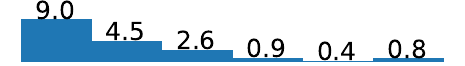} & 18.2 \\ \hline
RHC-S & \multicolumn{1}{c|}{\includegraphics[width=0.20\textwidth]{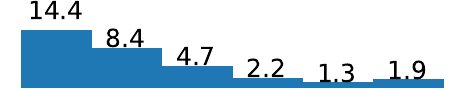}} & 32.8 & \multicolumn{1}{c|}{\includegraphics[width=0.20\textwidth]{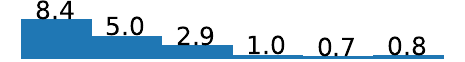}} & 18.7 & \includegraphics[width=0.20\textwidth]{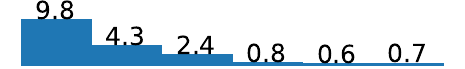} & 18.7 \\ \hline
SSRS & \multicolumn{1}{c|}{\includegraphics[width=0.20\textwidth]{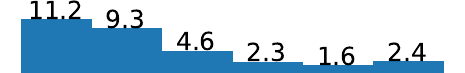}} & 31.4 & \multicolumn{1}{c|}{\includegraphics[width=0.20\textwidth]{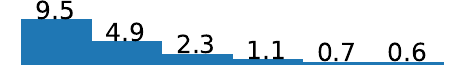}} & 19.1 & \includegraphics[width=0.20\textwidth]{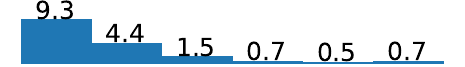} & 17.1 \\ \hline
SUPS & \multicolumn{1}{c|}{\includegraphics[width=0.20\textwidth]{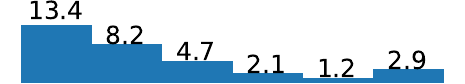}} & \textbf{32.7} & \multicolumn{1}{c|}{\includegraphics[width=0.20\textwidth]{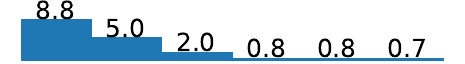}} & 18.1 & \includegraphics[width=0.20\textwidth]{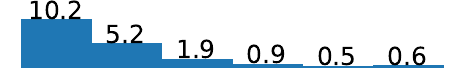} & 19.1\\ \hline
CES & \multicolumn{1}{c|}{\includegraphics[width=0.20\textwidth]{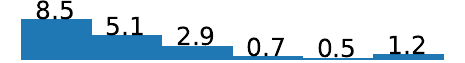}} & 18.9 & \multicolumn{1}{c|}{\includegraphics[width=0.20\textwidth]{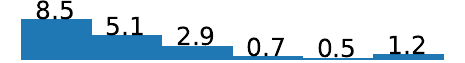}} & 18.9 & \includegraphics[width=0.20\textwidth]{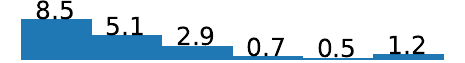} & 18.9 \\ \hline
SRS & \multicolumn{1}{c|}{\includegraphics[width=0.20\textwidth]{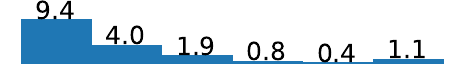}} & 17.7 & \multicolumn{1}{c|}{\includegraphics[width=0.20\textwidth]{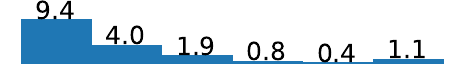}} & 17.7 & \includegraphics[width=0.20\textwidth]{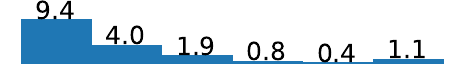} & 17.7
\\ \bottomrule
\end{tabular}}}
\end{table*}

\begin{table*}[t]
\centering
\caption{RQ2.2: 
 Average number of failures whose offset is
$ 12.5^\circ \leq \delta< 15^\circ$, $ 15^\circ \leq \delta< 17.5^\circ$, \dots $ 22.5^\circ \leq \delta< 25^\circ$. DD model}
\label{tab:fails_reg_DD}
\vspace{-6pt}
\renewcommand{\arraystretch}{1.4}
\footnotesize{
\resizebox{0.9\textwidth}{!}{
\begin{tabular}{c|cc|cc|c|c} \toprule
{Technique} &  \multicolumn{2}{c|}{LSA} & \multicolumn{2}{c|}{VAE} & \multicolumn{2}{c}{SAE}\\ \hline
GBS & \multicolumn{1}{c|}{\includegraphics[width=0.20\textwidth]{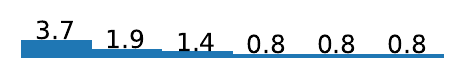}} & 9.4 & \multicolumn{1}{c|}{\includegraphics[width=0.20\textwidth]{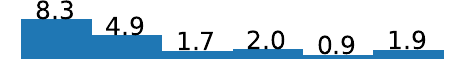}} & \textbf{19.7} & \includegraphics[width=0.20\textwidth]{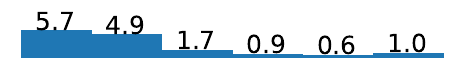} & 14.7 \\ \hline
DeepEST & \multicolumn{1}{c|}{\includegraphics[width=0.20\textwidth]{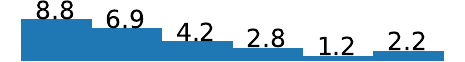}} & 26.1 & \multicolumn{1}{c|}{\includegraphics[width=0.20\textwidth]{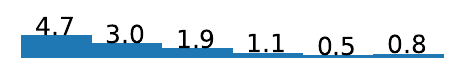}} & 12.0 & \includegraphics[width=0.20\textwidth]{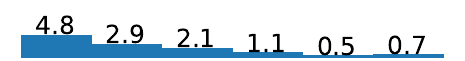} & 12.1 \\ \hline
2-UPS & \multicolumn{1}{c|}{\includegraphics[width=0.20\textwidth]{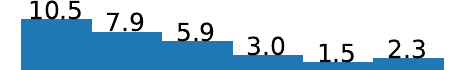}} & 31.1 & \multicolumn{1}{c|}{\includegraphics[width=0.20\textwidth]{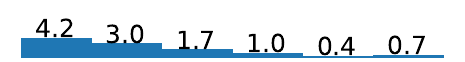}} & 11.0 & \includegraphics[width=0.20\textwidth]{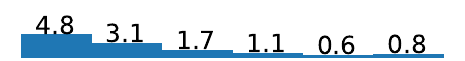} & 12.0 \\ \hline
RHC-S & \multicolumn{1}{c|}{\includegraphics[width=0.20\textwidth]{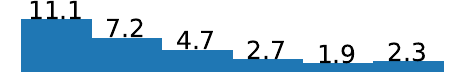}} & 29.9 & \multicolumn{1}{c|}{\includegraphics[width=0.20\textwidth]{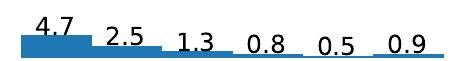}} & 10.6 & \includegraphics[width=0.20\textwidth]{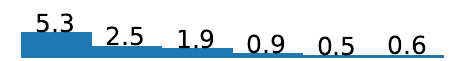} & 11.8 \\ \hline
SSRS & \multicolumn{1}{c|}{\includegraphics[width=0.20\textwidth]{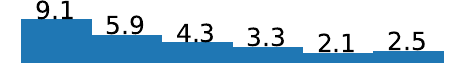}} & 27.3 & \multicolumn{1}{c|}{\includegraphics[width=0.20\textwidth]{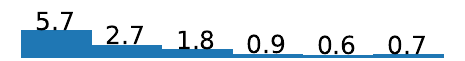}} & 12.4 & \includegraphics[width=0.20\textwidth]{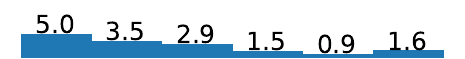} & \textbf{15.4} \\ \hline
SUPS & \multicolumn{1}{c|}{\includegraphics[width=0.20\textwidth]{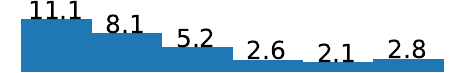}} & \textbf{31.9} & \multicolumn{1}{c|}{\includegraphics[width=0.20\textwidth]{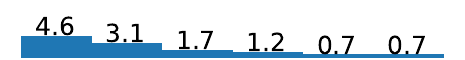}} & 12.1 & \includegraphics[width=0.20\textwidth]{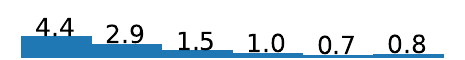} & 11.4 \\ \hline
CES & \multicolumn{1}{c|}{\includegraphics[width=0.20\textwidth]{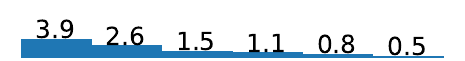}} & 10.3 & \multicolumn{1}{c|}{\includegraphics[width=0.20\textwidth]{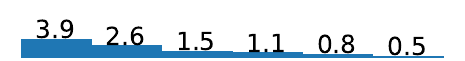}} & 10.3 & \includegraphics[width=0.20\textwidth]{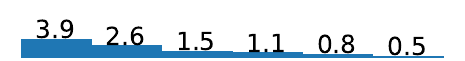} & 10.3 \\ \hline
SRS & \multicolumn{1}{c|}{\includegraphics[width=0.20\textwidth]{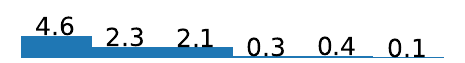}} & 9.8 & \multicolumn{1}{c|}{\includegraphics[width=0.20\textwidth]{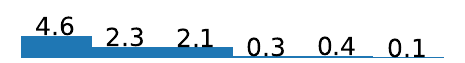}} & 9.8 & \includegraphics[width=0.20\textwidth]{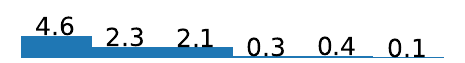} & 9.8
\\ \bottomrule
\end{tabular}}}
\vspace{-6pt}
\end{table*}

These results counterbalance the DeepEST and 2-UPS results on the estimates, which were worse than the others (RQ1.1).  DeepEST assumes that failures belong to a rare population, and is conceived to spot them. The greater ability to find misclassifications causes a greater variability of the estimates, and more budget is needed to converge. A similar problem is observed for 2-UPS. We hypothesize that partitioning combined with unequal sampling (both based on the auxiliary variable $\chi$) can push toward failing examples, but the estimator needs more time to converge. 
Unlike DeepEST, 2-UPS showed many spikes in the accuracy estimation; the estimator generates spikes every time a failure is detected with ``misleading'' values of $\chi$, namely misclassified examples with values of $\chi$ that would indicate a correct classification. For instance, failures with high \textit{confidence}, or with low LSA/DSA. SUPS and RHC-S seem very good compromises between the two -- more details in the final discussion.  
GBS, CES, and SRS detect fewer failures. For SRS and CES this is likely because the former does not use any auxiliary variable, the latter uses cross-entropy, not supposed to be related to failures. GBS and SSRS both use $\chi$ only for partitioning; but GBS detects fewer failures likely because the algorithm is thought to minimize the variance of the estimate. %

\subsubsection{RQ2.2: Regression}
Tables \ref{tab:fails_reg_DO} and \ref{tab:fails_reg_DD} report the histograms of the offset, starting from 12.5$^\circ$ to 25$^\circ$. 
Compared to the classification case, the differences here are less pronounced. Looking at the sum of the bins, we notice that CES and SRS select less examples with higher offset with respect to the others under the LSA case, while all the techniques are roughly equivalent with VAE and SAE\footnote{Note that the histograms for CES and SRS are the same along the three columns of the Table since they do not use LSA/VAE/SAE.}. GBS has similar poor performance, but it performs much better when used with VAE (consistently with the more unstable RMSE (Fig. \ref{fig:reg_bar_do}). SUPS is the best one with LSA. 
The good performance of  partitioning-based techniques (which achieve or even outperform DeepEST) is attributable to a better effect of partitioning when applied to regression compared to classification (since the auxiliary variable, used for partitioning, and the offset are more correlated). %

\subsection{RQ3: efficiency analysis}
\label{sec:sensitivity}

\subsubsection{RQ3.1. Accuracy assessment.} We synthesize in Tables \ref{tab:rq3_sens} and \ref{tab:rq3_sens_reg} the results for classification and regression. 
Besides the RMSE value at each point,\footnote{The full set of graphs for each dataset-auxiliary variable-model combination over the sample size are in the replication package.} we are interested in figuring out if the techniques smoothly converge as the sample size increase. First, we report for each dataset, technique, auxiliary variable, and model, how many times the minimum RMSE is reached under the given sample size. For instance, 3/3/3 of GBS for sample size 800 in MNIST, means that the minimum RMSE was reached for all the 3 models used with MNIST, using respectively \textit{confidence/LSA/DSA} as auxiliary variable. This is marked as green, and is the expected behaviour. When this is not true for at least one case, we mark it as red, and correspondingly mark as yellow those cells in the same row (with sample size smaller than 800) where the minimum was reached. 

\begin{table}[t]
\centering
\caption{RQ3.1: RMSE sensitivity analysis (conf./LSA/DSA)}
\label{tab:rq3_sens}
\setlength{\tabcolsep}{6pt}
\vspace{-6pt}
\resizebox{1\columnwidth}{!}{
\footnotesize{
\begin{tabular}{c|c|c|c|c|c|c|c} \toprule
  &  & \multicolumn{5}{c|}{{Minimum RMSE}} &  \multicolumn{1}{c}{{Inversions}}  \\
  & {Technique} & \textit{50} & \textit{100} & \textit{200} & \textit{400} & \textit{800} &  \small{\textit{{50}>{800}}} \\ \hline
   & GBS & 0/0/0 & 0/0/0 & 0/0/0 & 0/0/0 & \cellcolor[HTML]{34FF34}3/3/3 &  3/3/3 \\ \cline{2-8} 
   & DeepEST & 0/0/0 & 0/0/0 & 0/0/0 & \cellcolor[HTML]{FFFC9E}0/0/1 & \cellcolor[HTML]{FD6864}3/3/2 & 3/3/3 \\ \cline{2-8} 
   & 2-UPS & \cellcolor[HTML]{FFFC9E}1/0/0 & \cellcolor[HTML]{FFFC9E}1/0/0 & 0/0/0 & \cellcolor[HTML]{FFFC9E}0/1/0 & \cellcolor[HTML]{FD6864}1/2/3 &  \cellcolor[HTML]{FD6864}1/3/3 \\ \cline{2-8} 
   & RHC-S & 0/0/0 & 0/0/0 & 0/0/0 & \cellcolor[HTML]{FFFC9E}0/1/1 & \cellcolor[HTML]{FD6864}3/2/2 &  \cellcolor[HTML]{FFFFFF}3/3/3 \\ \cline{2-8} 
   & SSRS & 0/0/0 & 0/0/0 & 0/0/0 & \cellcolor[HTML]{FFFC9E}0/0/1 & \cellcolor[HTML]{FD6864}3/3/2 & 3/3/3 \\ \cline{2-8} 
   & SUPS & 0/0/0 & 0/0/0 & \cellcolor[HTML]{FFFC9E}1/0/0 & 0/0/0 & \cellcolor[HTML]{FD6864}2/3/3 & 3/3/3 \\ \cline{2-8} 
   & CES & 0 & 0 & 0 & 0 & \cellcolor[HTML]{34FF34}3 & 3 \\ \cline{2-8} 
  \multirow{-8}{*}{\rotatebox[origin=c]{90}{MNIST}} & SRS & 0 & 0 & 0 & 0 & \cellcolor[HTML]{34FF34}3 &  3 \\ \cmidrule{1-8}\morecmidrules\cmidrule{1-8} 
   & GBS & 0/0/0 & 0/0/0 & 0/0/0 & 0/0/0 & \cellcolor[HTML]{34FF34}3/3/3 &  3/3/3 \\ \cline{2-8} 
   & DeepEST & 0/0/0 & 0/0/0 & 0/0/0 & 0/0/0 & \cellcolor[HTML]{34FF34}3/3/3 & 3/3/3 \\ \cline{2-8} 
   & 2-UPS & 0/0/0 & \cellcolor[HTML]{FFFC9E}0/0/1 & 0/0/0 & \cellcolor[HTML]{FFFC9E}1/0/0 & \cellcolor[HTML]{FD6864}2/3/2  & 3/3/3 \\ \cline{2-8} 
   & RHC-S & 0/0/0 & 0/0/0 & 0/0/0 & \cellcolor[HTML]{FFFC9E}1/0/0 & \cellcolor[HTML]{FD6864}2/3/3 &  3/3/3 \\ \cline{2-8} 
   & SSRS & 0/0/0 & 0/0/0 & 0/0/0 & 0/0/0 & \cellcolor[HTML]{34FF34}3/3/3  & 3/3/3 \\ \cline{2-8} 
   & SUPS & 0/0/0 & 0/0/0 & 0/0/0 & \cellcolor[HTML]{FFFC9E}1/0/0 & \cellcolor[HTML]{FD6864}2/3/3 &  3/3/3 \\ \cline{2-8} 
   & CES & 0 & 0 & 0 & 0 & \cellcolor[HTML]{34FF34}3 &  3 \\ \cline{2-8} 
  \multirow{-8}{*}{\rotatebox[origin=c]{90}{CIFAR10}} & SRS & 0 & 0 & 0 & 0 & \cellcolor[HTML]{34FF34}3 & 3 \\ \cmidrule{1-8}\morecmidrules\cmidrule{1-8}  
   & GBS & 0/0/0 & 0/0/0 & 0/0/0 & \cellcolor[HTML]{FFFC9E}0/0/1 & \cellcolor[HTML]{FD6864}3/3/2 &  3/3/3 \\ \cline{2-8} 
   & DeepEST & 0/0/0 & 0/0/0 & 0/0/0 & 0/0/0 & \cellcolor[HTML]{34FF34}3/3/3 & 3/3/3 \\ \cline{2-8} 
   & 2-UPS & \cellcolor[HTML]{FFFC9E}0/0/1 & 0/0/0 & 0/0/0 & 0/0/0 & \cellcolor[HTML]{FD6864}3/3/2 &  \cellcolor[HTML]{FD6864}3/3/2 \\ \cline{2-8} 
   & RHC-S & 0/0/0 & 0/0/0 & 0/0/0 & \cellcolor[HTML]{FFFC9E}0/1/0 & \cellcolor[HTML]{FD6864}3/2/3  & 3/3/3 \\ \cline{2-8} 
   & SSRS & 0/0/0 & 0/0/0 & 0/0/0 & 0/0/0 & \cellcolor[HTML]{34FF34}3/3/3  & 3/3/3 \\ \cline{2-8} 
   & SUPS & 0/0/0 & 0/0/0 & 0/0/0 & \cellcolor[HTML]{FFFC9E}2/1/0 & \cellcolor[HTML]{FD6864}1/2/3 & \cellcolor[HTML]{FD6864}2/3/3 \\ \cline{2-8} 
   & CES & 0 & \cellcolor[HTML]{FFFC9E}1 & \cellcolor[HTML]{FFFC9E}1 & 0 & \cellcolor[HTML]{FD6864}1 &  \cellcolor[HTML]{FD6864}2 \\ \cline{2-8} 
\multirow{-8}{*}{\rotatebox[origin=c]{90}{CIFAR100}} & SRS & 0 & 0 & 0 & 0 & \cellcolor[HTML]{34FF34}3 & 3 \\ \bottomrule
\end{tabular}
}
}
\vspace{-6pt}
\end{table}

\begin{table}[t]
\centering
\caption{RQ3.1: RMSE sensitivity analysis (LSA/VAE/SAE)}
\label{tab:rq3_sens_reg}
\setlength{\tabcolsep}{5.4pt}
\vspace{-6pt}
\resizebox{1\columnwidth}{!}{
\small{
\begin{tabular}{c|c|c|c|c|c|c|c} \toprule
  &  & \multicolumn{5}{c|}{{Minimum RMSE}} &  \multicolumn{1}{c}{{Inversions}} \\
  & {Technique} & \textit{50} & \textit{100} & \textit{200} & \textit{400} & \textit{800} &  \small{\textit{{50}>{800}}} \\ \hline
   & GBS & 0/0/0 & 0/0/0 & \cellcolor[HTML]{FFFC9E}0/2/0 & \cellcolor[HTML]{FFFC9E}0/0/1 & \cellcolor[HTML]{FD6864}2/0/1 &  2/2/2 \\ \cline{2-8} 
   & DeepEST & 0/0/0 & 0/0/0 & 0/0/0 & 0/0/0 & \cellcolor[HTML]{34FF34} 2/2/2 & 2/2/2 \\ \cline{2-8} 
   & 2-UPS & 0/0/0 & 0/0/0 & 0/0/0 & \cellcolor[HTML]{FFFC9E}1/1/0 & \cellcolor[HTML]{FD6864}1/1/2 &  2/2/2 \\ \cline{2-8} 
   & RHC-S & 0/0/0 & 0/0/0 & 0/0/0 & 0/0/0 & \cellcolor[HTML]{34FF34}2/2/2  & 2/2/2 \\ \cline{2-8} 
   & SSRS & 0/0/0 & 0/0/0 & 0/0/0 & 0/0/0 & \cellcolor[HTML]{34FF34}2/2/2 & 2/2/2 \\ \cline{2-8} 
   & SUPS & 0/0/0 & 0/0/0 & 0/0/0 & \cellcolor[HTML]{FFFC9E}1/0/0 & \cellcolor[HTML]{FD6864}1/2/2 &  2/2/2 \\ \cline{2-8} 
   & CES & 0 & 0 & \cellcolor[HTML]{FFFC9E}1 & 0 & \cellcolor[HTML]{FD6864}1 &  2 \\ 
   \cline{2-8} 
\multirow{-7}{*}{\rotatebox[origin=c]{90}{Udacity}} & SRS & 0 & 0 & 0 & 0 & \cellcolor[HTML]{34FF34}2 &  2 \\ \bottomrule
\end{tabular}%
}
}
\vspace{-6pt}
\end{table}

\begin{table}[t]
 \centering
 \caption{RQ3.2: Failures sensitivity analysis (conf./LSA/DSA)}
 \label{tab:rq3_sens_fails}
 \vspace{-6pt}
\setlength{\tabcolsep}{5.5pt}
 \resizebox{.97\columnwidth}{!}{
 \small{
 \begin{tabular}{c|c|c|c|c} \toprule
  \textit{} & Technique & \textit{mean(min)} & \textit{$F_{800/50}$} & \textit{mean(max)} \\ \hline
  \multirow{8}{*}{\rotatebox[origin=c]{90}{MNIST}} & GBS & 5.3/5.8/4.5 & 26.3/15.7/21.9 & 136.5/91.9/95.9 \\ \cline{2-5} 
   & DeepEST & 19.3/8.0/17.9 & 16.7/17.7/16.5 & 321.6/140.0/295.9 \\ \cline{2-5} 
   & 2-UPS & 17.4/5.8/5.9 & 15.9/17.1/16.9 & 277.2/100.5/98.0 \\ \cline{2-5} 
   & RHC-S & 17.1/6.7/6.2 & 16.1/15.1/16.0 & 274.2/101.8/100.3 \\ \cline{2-5} 
   & SSRS & 10.0/6.9/5.5 & 15.5/15.8/16.6 & 153.3/107.4/90.9 \\ \cline{2-5} 
   & SUPS & 17.7/6.2/5.8 & 16.0/16.4/17.3 & 282.6/102.6/100.9 \\ \cline{2-5} 
   & CES & 3.8 & 16.1 & 61.5 \\ \cline{2-5} 
   & SRS & 3.8 & 15.9 & 58.6 \\ \cmidrule{1-5}\morecmidrules\cmidrule{1-5} 
  \multirow{8}{*}{\rotatebox[origin=c]{90}{CIFAR10}} & GBS & 15.4/14.9/14.4 & 17.9/15.9/18.3 & 272.5/238.7/261.1 \\ \cline{2-5} 
   & DeepEST & 26.9/17.1/25.3 & 16.1/16.2/16.2 & 432.7/277.7/408.6 \\ \cline{2-5} 
   & 2-UPS & 26.7/15.8/16.8 & 16.1/16.1/15.8 & 431.8/252.7/264.2 \\\cline{2-5} 
   & RHC-S & 27.2/15.9/16.1 & 15.7/16.0/16.3 & 427.7/252.6/262.4 \\ \cline{2-5} 
   & SSRS & 19.7/15.3/14.4 & 16.0/15.7/16.0 & 315.1/239.4/231.1 \\ \cline{2-5} 
   & SUPS & 26.7/15.1/16.4 & 16.2/16.6/16.2 & 433.8/250.6/263.3 \\ \cline{2-5} 
   & CES & 14.1 & 15.2 & 216.5 \\ \cline{2-5} 
   & SRS & 13.8 & 16.4 & 227.6 \\ \cmidrule{1-5}\morecmidrules\cmidrule{1-5} 
  \multirow{8}{*}{\rotatebox[origin=c]{90}{CIFAR100}} & GBS & 21.5/21.3/21.7 & 15.9/15.9/16.5 & 341.0/339.2/357.8 \\ \cline{2-5} 
   & DeepEST & 33.7/29.8/34.9 & 16.2/16.0/11.3 & 546.6/475.3/393.3 \\ \cline{2-5} 
   & 2-UPS & 35.7/27.2/24.5 & 15.9/16.0/15.7 & 565.9/434.7/385.8 \\ \cline{2-5} 
   & RHC-S & 35.3/27.6/24.1 & 15.9/15.7/15.9 & 560.9/432.6/383.3 \\ \cline{2-5} 
   & SSRS & 28.2/22.7/20.8 & 15.5/16.5/15.9 & 436.5/374.5/329.8 \\ \cline{2-5} 
   & SUPS & 35.2/27.4/23.8 & 16.2/16.2/16.4 & 571.2/441.1/389.7 \\ \cline{2-5} 
   & CES & 19.7 & 13.8 & 270.5 \\ \cline{2-5} 
   & SRS & 20.8 & 15.2 & 315.0 \\ \bottomrule
 \end{tabular}
 }
 }
\end{table}

There are many cases where the minimum is not achieved with the largest sample size (red cells). For instance, the instability of 2-UPS makes it even reach the best values with a sample size 50 (MNIST and CIFAR100) and sample size 100 (MNIST and CIFAR10). %
CES with CIFAR100 has the same convergence problem, while it is stable in MNIST and CIFAR10. In remaining red cases, the minimum is at 400. 
SRS is the most stable technique, for  independence from auxiliary variables. 
GBS and DeepEST are stable for 2 of 3 datasets; in the bad case, they converge at size 400. %
For regression, performance is better; 
GBS is more unstable, while the others converge at 800 with few exceptions at 400 and one (CES) at 200.

Tables \ref{tab:rq3_sens} and \ref{tab:rq3_sens_reg} report also in how many cases the RMSE with budget $50$ is smaller than that with budget $800$ (red cells). We call this  \textit{inversions}, %
denoting convergence problems. %
There are 5 such cases:
3 for 2-UPS (2 with MNIST and 1 with CIFAR100), 1 for SUPS and 1 for CES (both with CIFAR100). Inversions never occur for regression. %
SRS is still the most stable technique for both classification and regression.
Again, 2-UPS is the most affected one.

\subsubsection{RQ3.2: Failing examples detection.} Results for this RQ are in Tables \ref{tab:rq3_sens_fails} and \ref{tab:rq3_sens_fails_reg}. 
For regression, we consider as failures all the predictions with an error on the steering angle greater than $12.5^{\circ}$. 
The Table reports the mean (over the sample sizes) of the minimum and maximum number of failures, and
the ratio between the number of failures detected with sizes 800 and 50 ($F_{800/50}$). 

We observe there is no \textit{inversion}: failures constantly increase with the budget size -- they roughly double as the sample size doubles for both classification and regression (detailed values are in the replication package). The expectation in this case is fully matched by all techniques. Looking at $F_{800/50}$, all the results of RQ2 are confirmed for all budget sizes.

 \begin{table}[t]
 \centering
 \caption{RQ3.2: Failures sensitivity analysis (LSA/VAE/SAE)}
 \label{tab:rq3_sens_fails_reg}
 \setlength{\tabcolsep}{5.4pt}
 \vspace{-6pt}
 \resizebox{.99\columnwidth}{!}{
 \small{
 \begin{tabular}{c|c|c|c|c} \toprule
  \textit{} & Technique & \textit{mean(min)} & \textit{$F_{800/50}$} & \textit{mean(max)} \\ \hline
 \multirow{8}{*}{\rotatebox[origin=c]{90}{Udacity}} & GBS & 4.2/3.7/4.1 & 3.0/34.1/14.9 & 12.4/122.8/57.8 \\ \cline{2-5} 
  & DeepEST & 6.8/3.6/3.6 & 16.2/16.6/18.0 & 109.3/60.3/61.2 \\ \cline{2-5} 
  & 2-UPS & 7.4/3.8/3.4 & 16.6/15.5/18.9 & 123.1/59.1/61.2 \\ \cline{2-5} 
  & RHC-S & 8.1/3.5/3.7 & 13.8/17.3/16.4 & 112.1/60.8/61.5 \\ \cline{2-5} 
  & SSRS & 7.5/4.2/4.2 & 15.6/15.1/15.2 & 117.5/64.2/64.9 \\ \cline{2-5} 
  & SUPS & 8.1/3.7/3.8 & 15.9/16.5/16.2 & 127.3/60.6/61.1 \\ \cline{2-5} 
  & CES & 3.1 & 20.9 & 65.3 \\ \cline{2-5} 
  & SRS & 3.3 & 17.9 & 58.3 \\ \bottomrule
 \end{tabular}
 }
 }
\vspace{-6pt}
 \end{table}

\section{Discussion }
\label{sect:Discussion}

We analyze the results with respect to the main impacting factors, to provide guidance to both practitioners (to select the technique best fitting the needs) and researchers (to design new  techniques).

The performance of a sampling technique depends on the tester's \textit{objective} and on the application \textit{context}. 

As for the \textit{objective} (set in the problem formulation, Sec. \ref{sect:DeepSample_form}),
while a tester is always interested in an \textit{unbiased assessment} of the DNN accuracy, s/he can specifically focus on: 
\begin{itemize}
    \item[\textcircled{\raisebox{-.75pt}1}]
\textit{High confidence (i.e., low variance)}, e.g., as criterion to release a DNN, or to choose which DNN to deploy among various alternatives -- a high-confidence estimate is usually required in critical domains. This can be achieved by reducing the RMSE or RMedSE: in the former case, one looks for high-confidence estimate even in presence of outliers; 
 in the latter case, one neglects the negative effect of outliers. 
    \item[\textcircled{\raisebox{-.75pt}2}]
 \textit{High failure exposure ability}, e.g., when the tester needs to assess and improve the DNN accuracy efficiently, and the high-confidence requirements can be relaxed (e.g., in non-critical domains). The simultaneous assessment and improvement can help during subsequent re-training/fine-tuning iterations to efficiently track progress in the achieved accuracy. 
    \item[\textcircled{\raisebox{-.75pt}3}]
 \textit{A trade-off between confidence in the accuracy estimate and number of exposed failures}, e.g., 
 when a good confidence estimate is used to monitor the accuracy of a DNN and engineers want to use the exposed failing examples in the re-training actions (these may be triggered only when the accuracy drops under a certain threshold) \cite{Guerriero23_nier}.
\end{itemize}

As for the \textit{context}, following our experimental design, the factors that we identified as potentially impacting are: the \textbf{task} (classification or regression), the \textbf{sample size} (hence the budget available), the \textbf{dataset}\footnote{Datasets and models are considered together; the average accuracy of the models  on the datasets capture three distinct cases of low, %
medium
and high 
accuracy (Tab. \ref{tab:models})}, 
and the \textbf{auxiliary variable}, if available, for sampling.

\begin{table}[t]
\centering
\caption{Top 3 techniques across configurations (two factors) - (T: trade-off, nf: number of failures)}%
\label{tab:top3:two_way_objectives}
\vspace{-10pt}
\footnotesize{
\subfloat[Task]{
\begin{tabular}{r|c|c|c||c|c|c}
\toprule 
&\multicolumn{3}{c||}{Classification} & \multicolumn{3}{c}{Regression}\\\cline{1-4}\cline{5-7}

& 1$^{st}$ & 2$^{nd}$ & 3$^{rd}$ & 1$^{st}$ & 2$^{nd}$ & 3$^{rd}$\\\cline{2-4}\cline{5-7}

{RMSE}& GBS &SSRS &SUPS &CES &SSRS &RHC-S \\\cline{2-4}\cline{5-7}
{RMedSE}&  SSRS &GBS &SRS & SSRS &SUPS &RHC-S\\\cline{2-4}\cline{5-7}
{Failures} & DeepEST &SUPS &RHC-S & SSRS &SUPS &GBS\\\cline{2-4}\cline{5-7}
T$_\textrm{RMSE-nf}$& SUPS &RHC-S &DeepEST & SSRS &SUPS&RHC-S\\\cline{2-4}\cline{5-7}
T$_\textrm{RMedSE-nf}$& SUPS &RHC-S &DeepEST & SSRS &SUPS&RHC-S\\
\bottomrule
\end{tabular}
\vspace{-5pt}
}
\\ \vspace{-6pt}
\subfloat[Sample size]{
\begin{tabular}{r|c|c|c||c|c|c}
\toprule 
&\multicolumn{3}{c||}{Small/Medium (50, 100, 200)} & \multicolumn{3}{c}{Large (400, 800)}\\\cline{1-4}\cline{5-7}
& 1$^{st}$ & 2$^{nd}$ & 3$^{rd}$ & 1$^{st}$ & 2$^{nd}$ & 3$^{rd}$\\\cline{2-4}\cline{5-7}
{RMSE}& SSRS & GBS & SUPS &GBS &SSRS &SRS \\\cline{2-4}\cline{5-7}
{RMedSE}& SSRS &SRS&SUPS & SSRS &GBS &SRS\\\cline{2-4}\cline{5-7}
{Failures}& DeepEST &RHC-S &SUPS & SUPS &DeepEST &2-UPS\\\cline{2-4}\cline{5-7}
T$_\textrm{RMSE-nf}$& RHC-S &SUPS &SSRS & SUPS &RHC-S&GBS\\\cline{2-4}\cline{5-7}
T$_\textrm{RMedSE-nf}$& RHC-S &SUPS &SSRS & SUPS &RHC-S&SSRS\\
\bottomrule
\end{tabular}
\vspace{-5pt}
}
\\ \vspace{-6pt}
\subfloat[Dataset/model accuracy]{
\begin{tabular}{r|c|c|c||c|c|c}
\toprule 
&\multicolumn{3}{c||}{Low (CIFAR10, CIFAR100)} & \multicolumn{3}{c}{High (MNIST, Udacity)}\\\cline{1-4}\cline{5-7}

& 1$^{st}$ & 2$^{nd}$ & 3$^{rd}$ & 1$^{st}$ & 2$^{nd}$ & 3$^{rd}$\\\cline{2-4}\cline{5-7}

{RMSE}& GBS &SSRS &SRS & SSRS&RHC-S & SUPS \\\cline{2-4}\cline{5-7}
{RMedSE}&SSRS &GBS &SRS & SSRS& RHC-S & SUPS\\\cline{2-4}\cline{5-7}
{Failures}& RHC-S &2-UPS&DeepEST & DeepEST&SUPS & RHC-S\\\cline{2-4}\cline{5-7}
T$_\textrm{RMSE-nf}$& SUPS &RHC-S&GBS & SUPS & SSRS & RHC-S\\\cline{2-4}\cline{5-7}
T$_\textrm{RMedSE-nf}$&SUPS &RHC-S&GBS/2-UPS & SUPS & SSRS & RHC-S\\
\bottomrule
\end{tabular}
}
\\ \vspace{-6pt}
\subfloat[Auxiliary variable]{
\begin{tabular}{c|c|c|c|c|c}
\toprule
& {RMSE} & {RMedSE} & {Failures} & T$_\textrm{RMSE-nf}$ & T$_\textrm{RMedSE-nf}$ \\\hline
\multirow{3}{*}{conf.} &1$^{st}$: SSRS  &SSRS &SUPS  &SUPS  &SUPS  \\
 &2$^{nd}$: GBS  &GBS  &2-UPS  &RHC-S  &RHC-S  \\
 &3$^{rd}$: SRS  &SRS  &RHC-S  &2-UPS  &2-UPS  \\\hline
\multirow{3}{*}{\begin{tabular}[c]{@{}c@{}}LSA \\ $_\textrm{classification}$\end{tabular}} & GBS &SSRS  &DeepEST  & RHC-S & RHC-S \\
& SUPS&GBS  & RHC-S &SUPS  &SUPS    \\
& RHC-S&SRS  &SUPS  & DeepEST & DeepEST   \\\hline

\multirow{3}{*}{DSA} &SUPS  &SUPS  &DeepEST  & SUPS &SUPS  \\
 &GBS  &GBS  &2-UPS  &RHC-S  &RHC-S  \\
 &SSRS  &SSRS  &SUPS  &DeepEST  &DeepEST  \\\hline

\multirow{3}{*}{\begin{tabular}[c]{@{}c@{}}LSA \\ $_\textrm{regression}$\end{tabular}} &SSRS  & SSRS &  SSRS&  SSRS& SSRS \\
 &RHC-S  &SUPS  &SUPS  & RHC-S &SUPS  \\
 &SUPS  &RHC-S  & 2-UPS &SUPS  &RHC-S  \\\hline

\multirow{3}{*}{VAE} &CES  & SRS &GBS  & SSRS & SSRS \\
 &SRS  &SSRS  &SSRS  & CES &SUPS  \\
 &RHC-S  &RHC-S  &SUPS  & SUPS &GBS  \\\hline

\multirow{3}{*}{SAE} &SRS  & SSRS &GBS  & CES & SSRS \\
 &CES  &RHC-S  &DeepEST  &SRS  &RHC-S  \\
 &SSRS  & CES & RHC-S & RHC-S &SUPS  \\

\bottomrule
\end{tabular}
}
}
\vspace{-3pt}
\end{table}

Table \ref{tab:top3:two_way_objectives} reports a two-way analysis of the ranking performance of the techniques. %
 On the row, we list the objective. On the column, we break down the results by the impacting factor. 
 For each combination (e.g, RMSE with Classification, Table \ref{tab:top3:two_way_objectives}a),  we count the number of times a technique was among the top-3 ones, 
 and report the best 3 techniques according to this count. 
 
 A practitioner should consider the combination reflecting more his/her needs and context. For instance, one might want a high-confidence robust-to-outlier assessment (row 1), with a medium (200) labelling effort (Table \ref{tab:top3:two_way_objectives}b); or (s)he might not want to use LSA or DSA, which are more expensive to compute, preferring the use of confidence (Table \ref{tab:top3:two_way_objectives}d).\footnote{These results have to be read with the pairwise statistical test results, as the best 3 techniques could be negligibly different (in which case one can be chosen arbitrarily).} 
 Since exploring any \textit{n-}way combinations could be of interest too (e.g., small RMSE \textit{and}  small sample size \textit{and} high-accuracy dataset), we release a notebook in our replication package\cref{repo} to specify the factors of interest and query the results. %

\begin{table}[t]
\centering
\caption{Number of best-performing occurrences out of 270 (classification) and 60 (regression) configurations}
\label{tab:top}
\vspace{-6pt}
\resizebox{1.\columnwidth}{!}{
\small{
\begin{tabular}{c|c|c|c||c|c|c|c}
\toprule
\textit{} & \multicolumn{3}{c||}{{Classification}} & \textit{} & \multicolumn{3}{c}{{Regression}} \\ \hline
aux. & \multicolumn{1}{c|}{{RMSE}} & RMedSE & {Failures} & aux. & \multicolumn{1}{c|}{{RMSE}} & RMedSE & {Failures} \\ \hline
conf. & \multicolumn{1}{c|}{73} & 82 & {243} & LSA & \multicolumn{1}{c|}{{35}} & 30 & {52} \\ \hline
LSA & \multicolumn{1}{c|}{{85}} & 82 & 8 & VAE & \multicolumn{1}{c|}{12} & 17 & {7} \\ \hline
DSA & \multicolumn{1}{c|}{{112}} & 106 & {19} & SAE & \multicolumn{1}{c|}{{13}} & 13 & 1 \\ \bottomrule
\end{tabular}
}
}
\end{table}

Besides combination-specific findings easily inferable from the Tables, some interesting patterns are hereafter highlighted:
\begin{itemize}%
\item In high-confidence assessment \textcircled{\raisebox{-.75pt}1} (RMSE, RMedSE), SSRS is among the best three techniques in 22 out of 24 combinations, followed by GBS and SRS (12/24), SUPS (11/24) and RHC-S (10/24).  %
The existing techniques CES and DeepEST are never in the top 3. Surprisingly, SRS appears often, especially for large sample size, and for low-accuracy models; %
\item In high failure exposure \textcircled{\raisebox{-.75pt}2}, SUPS stands out 10 out of 12 times, followed by DeepEST (8/12) and RHC-S (7/12);
\item For good trade-offs \textcircled{\raisebox{-.75pt}3} ($T_{RMSE\text{-}nf}$, $T_{RMedSE\text{-}nf}$), SUPS and RHC-S appear almost always 
(23/24 and 22/24, respectively). The others are far less common (SSRS 12/24, DeepEST 6/24).
\end{itemize}

It is worth to note that the new algorithms proposed (GBS, 2-UPS, RHC-S, SSRS, SUPS) appear among the best three in the vast majority of cases. The following specific considerations can be drawn.

SSRS is particularly good for high-confidence estimates; SUPS (and to a lesser extent RHC-S) outperforms the others for high failure exposure, where it even defeats DeepEST that is specifically conceived for that task via adaptive sampling.

SUPS and RHC-S give the best trade-offs. This indicates that they perform generally well for all the objectives.

The distinguishing feature of the new techniques is that they exploit the auxiliary variable for just partitioning (SSRS, GBS) and/or for inputs selection (RHC-S, SUPS, 2-UPS). This in essence allows to direct the sampling toward higher-variance areas of the population, 
reducing the estimator variance and exposing more failures.

\vspace{3pt}
In the perspective of a researcher devising a new technique, attention has to be paid to these aspects: auxiliary variable (if and which one to use), partitioning, and replacement scheme (Tab. \ref{summary:table}).

\vspace{3pt}
\noindent\textbf{Auxiliary variable}. The performance of auxiliary variables is useful not only for selecting a technique, but also to design new ones. %
The results in Table \ref{tab:top3:two_way_objectives}.d highlight that the only techniques not using auxiliary variables (SRS and CES) are rarely among the top-3 ones, especially for the failure exposure ability ($A_f$). %

Table \ref{tab:top} reports how many times each auxiliary variable yields the best RMSE and RMedSE, and the number of failures.  
For classification, DSA and \textit{confidence} are the best variables for RMSE/RMedSE \textcircled{\raisebox{-.75pt}1} and number of failures \textcircled{\raisebox{-.75pt}2}, respectively. 
It is important to highlight that \textit{confidence} is cheaper to collect, as it comes with the output of the classification. %
For regression, LSA shows the best results \textcircled{\raisebox{-.75pt}1}\textcircled{\raisebox{-.75pt}2}\textcircled{\raisebox{-.75pt}3}. The variables derived by SAE/VAE perform poorly.  %

\vspace{3pt}
\noindent \textbf{Partitioning}. Partitioning based on auxiliary variables is particuarly beneficial for good accuracy estimates \textcircled{\raisebox{-.75pt}1}; SSRS and GBS are the best ones for this aim. %
The benefit of partitioning is lower when the aim is to expose 
failures \textcircled {\raisebox{-.75pt}2}\textcircled{\raisebox{-.75pt}3}; performance is better when partitioning 
with LSA, especially for regression, as it is better correlated to (in)accuracy. 

\vspace{3pt}
\noindent
\textbf{Replacement}. We found no remarkable advantage
 of without-replacement sampling; for instance, 
 SUPS (with replacement) works well %
 in all scenarios. %
This is likely due to the negligible sample size compared to the operational dataset, hence sampling with replacement is unlikely to pick the same example twice.

\section{Threats to validity}
\label{sect:Threats}
As for the selection of the experimental subjects, we have considered publicly available DNNs \cite{Guerriero23}; we have however re-trained them from scratch to have realistic accuracy and avoid the mentioned inflated accuracy issue described in \cite{Recht19}. %

The choice of the sample size affects the results. We ran a sensitivity analysis with five (from 50 to 800) values of the sample size. Different values could yield different results. 

The evaluation does not include an extensive analysis of partitioning. We ran $k$-means, %
with $k=10$ partitions, after a preliminary tuning on 30 random samples from MNIST and $k= 6, 8, 10, 12$. Extending the tuning of $k$ to all cases would improve performance. 

Despite extensive code inspection, 
the presence of defects in the algorithms cannot be excluded. 

External validity is undermined by the number of models and datasets; we considered state-of-the-art DNNs and widely-used datasets. %
The replicability of the experiments 
mitigates this threat.

\section{Conclusions}
\label{sect:Conclusions}
We presented \texttt{DeepSample}, a framework encompassing a set of sampling-based techniques for DNN operational accuracy assessment. We implemented techniques with and without partitioning, with and without replacement, with and without auxiliary variables to drive the selection, and we empirically evaluated them in terms of accuracy estimation and number of failures,
on both classification and regression problems. 

The findings pertaining to the individual techniques, as well as to the key factors impacting the sampling algorithms, serve: \textit{i)} as guidance for testers to select the technique depending on the needs 
and on the auxiliary information available to expedite sampling, and \textit{ii)} for researchers to devise new techniques. %

We conclude that the tester's objective and the application context are crucial in selecting a sampling technique. Techniques yielding high-confidence estimates (such as SSRS) are well suited to check the DNN against a release criterion, or for choosing among different DNNs. Techniques with high failure exposure ability (such as SUPS and DeepEST) are well suited for the simultaneous DNN accuracy assessment and improvement in iterative life cycle models. Techniques exhibiting a good trade-off between high-confidence estimates and high failure exposure (such as SUPS and RHC-S) are appropriate for cost-effective assessment and retraining. 

In devising new techniques, the use of auxiliary variables and partitioning is strongly encouraged, as they have been shown to be beneficial for both accuracy estimation and failure exposure -- LSA was the best choice for regression, while confidence (for failures exposure) and DSA (for accuracy estimation) were the best ones for classification. 

\section{Data Availability}
All results and the artefacts for replication are available at: \\
\url{https://github.com/dessertlab/DeepSample.git}.

\section*{Acknowledgment}
This project has received funding from the European Union's Horizon 2020 research and innovation programme under the Marie Sk{\l}odowska-Curie grant agreement No 871342 “uDEVOPS”.

\bibliographystyle{unsrt}

\end{document}